\documentclass{jfm}
\usepackage{graphicx}
\usepackage{epstopdf, epsfig}
\usepackage{subfig}
\usepackage{epsf}
\usepackage{color}
\usepackage{colortbl}
\usepackage{array}
\usepackage{booktabs}
\usepackage{color}
\usepackage{bm}
\usepackage{hyperref}
\usepackage{natbib}
\usepackage{amsmath}
\usepackage{array}
\usepackage{upgreek}
\usepackage[dvipsnames]{xcolor}
\hypersetup{
    colorlinks = true,
    urlcolor   = black,
    citecolor  = black,
}

\newcommand{\RomanNumeralCaps}[1]
\linenumbers
\newcommand{\comm}[1]{}

\definecolor{Reviewer1Color}{RGB}{255, 0, 0}    
\definecolor{Reviewer2Color}{RGB}{255,0,255}    
\definecolor{Reviewer3Color}{RGB}{0, 0, 255}    
\definecolor{ReviewercollectiveColor}{RGB}{255, 140, 0}    

\newcommand{\rea}[1]{\textcolor{Reviewer1Color}{#1}}
\newcommand{\reb}[1]{\textcolor{Reviewer2Color}{#1}}
\newcommand{\rec}[1]{\textcolor{Reviewer3Color}{#1}}
\newcommand{\red}[1]{\textcolor{ReviewercollectiveColor}{#1}}

\title{Wake transitions and melting dynamics of a translating sphere in warm liquid}

\author{Zhong-Han Xue \and Jie Zhang\corresp{\email{j\_zhang@xjtu.edu.cn}}}

\affiliation{State Key Laboratory for Strength and Vibration of Mechanical Structures, School of Aerospace, Xi'an Jiaotong University, Xi'an, Shaanxi 710049, PR China}

\begin{document}
\maketitle

\begin{abstract}
We investigate the three-dimensional melting dynamics of an initially spherical particle translating in a warmer liquid using sharp-interface simulations that fully resolve both solid and fluid phases with the Stefan condition. A wide parameter space is explored, spanning initial Reynolds number ($\mathit{Re}_0$), Stefan number ($\mathit{St}$), and Richardson number ($\mathit{Ri}$). In the absence of buoyancy (\rea{$\mathit{Ri}= 0$}), the interface evolution is governed by canonical wake bifurcations. Four regimes are identified: an axi-symmetric regime ($\mathit{Re}_0<212$) with a rounded front and planar rear; a steady-planar-symmetric regime ($212<\mathit{Re}_0<273$) with an inclined rear plane; a periodic-planar-symmetric regime ($273<\mathit{Re}_0<355$) where vortex shedding emerges in the wake; and a chaotic regime ($\mathit{Re}_0>355$) with fluctuating stagnation points and a more rounded rear. Despite these differences, all regimes exhibit a tendency toward melt-rate homogenisation over time. Besides, we introduce an aspect-ratio-based surface-area formulation that yields a predictive model, accurately capturing volume evolution across regimes. Hydrodynamic loads also reflect the coupling between shape and flow: drag follows rigid-sphere correlations only at moderate $\mathit{Re}_0$; planar rears enhance drag at higher $\mathit{Re}_0$; lift appears only in symmetry-broken regimes and reverses late in time; torque reorients the rear plane toward vertical, consistent with free-body experiments. When buoyancy is included, assisting configurations ($\mathit{Ri}>0$) suppress recirculation and maintain quasi-spherical shapes, whereas opposing or transverse buoyancy ($\mathit{Ri}<0$) destabilises wakes and promotes tilted planar rears. These results provide a unified framework for convection-driven melting across laminar, periodic, and chaotic wakes, with implications for geophysical and industrial processes.
\end{abstract}


\section{Introduction}\label{sec:introduction}

Melting of ice bodies or solid particles plays a central role in diverse natural and industrial contexts, ranging from oceanography and planetary science to astrophysics and metallurgy \citep{davis2001theory}. When external flows are present, whether driven by buoyancy or imposed by forced convection, the surrounding temperature field is altered, producing complex body interface morphologies during melting. These evolving interfaces, in turn, modify the flow by acting as moving boundaries. Understanding this coupled fluid-thermal-interface dynamics is essential for predicting iceberg ablation \citep{cenedese2023icebergs}, interpreting the formation of magma oceans \citep{ulvrova2012numerical}, reconstructing the thermal histories of icy moons \citep{gastine2024rotating}, and controlling solidification defects in metallurgical processes \citep{shevchenko2013chimney}.

Over the past two decades, significant progress has been achieved through laboratory experiments, numerical simulations, and theoretical analyses. Much of this progress has relied on simplified model systems. A canonical configuration is Rayleigh–Bénard convection \citep{lohse2024ultimate}, where buoyancy drives turbulent motion that interacts with a melting boundary. The closed geometry and high experimental controllability of this setup make it an attractive framework for studying melting dynamics \citep{du2024physics}. Foundational contributions include the experiments by \citet{davis1984pattern} and the simulations by \citet{ulvrova2012numerical} \citet{rabbanipour2018basal}, and \citet{favier2019rayleigh}, which revealed unsteady processes such as convective roll mergers and shell-like melting fronts. These studies also proposed scaling laws for interface evolution and melting rates derived from energy balances. Recent extensions of this framework have incorporated additional physical complexities to better approximate geophysical and engineering scenarios. Examples include the anomalous temperature dependence of water density \citep{wang2021growth}, ambient shear and turbulence \citep{couston2021topography, ravichandran2022combined, yang2023morphology}, and double-diffusive effects in solute-laden systems \citep{xue2024flow, guo2025effects}. Such modifications have been shown to exert profound influence on interface morphology and melting rates. For instance, \citet{couston2021topography} demonstrated that turbulent kinetic energy promotes ripple- and scallop-like structures, while \citet{xue2024flow} showed that solute concentration gradients can suppress convective transport, thereby inhibiting morphological pattern formation.

In contrast to Rayleigh–Bénard convection, where melting typically occurs at extended boundaries, a complementary class of problems involves melting bodies of finite extent immersed in open flows. This scenario is particularly relevant to iceberg ablation, which exhibits nonlinear sensitivity to ocean currents \citep{fitzmaurice2017nonlinear}. Both the magnitude and shear of surface flows strongly control melting rates and interface morphology \citep{fitzmaurice2018parameterizing, wagner2014footloose}. A simplified yet fundamental model considers a stationary ice body in a uniform current, isolating the role of forced convection from buoyancy-driven drift. Pioneering experiments by \citet{hao2001melting, hao2002heat} on ice spheres in horizontal flows demonstrated that melting rates increase with both fluid velocity and temperature, and that enhanced wake circulation downstream promotes heat transfer, leading to flattened rear surfaces. Orientation-dependent melting rates have since been widely reported. For example, \citet{hester2021aspect} showed that vertical faces melt faster than horizontal ones in saline flows. More recently, numerical studies by \cite{yang2024shape, yang2024circular} and \cite{xu2025aspect} established that melting rate depends non-monotonically on body aspect ratio, and developed scaling arguments based on distinct conductive and convective contributions. Despite such complexities, experiments and simulations consistently indicate that the upstream face of melting spheres converges to a self-similar profile \citep{hao2001melting, yang2024shape}. Comparable shape evolution has also been reported for eroding or dissolving bodies in unidirectional flows \citep{moore2013self, mac2015shape, ristroph2012sculpting, huang2020ultra}, where boundary-layer models accurately capture the emergence of self-similar forms. In turbulent environments, generated for instance by meltwater plumes or subglacial discharge, local flow intensification can further accelerate melting rates \citep{machicoane2013melting, mccutchan2024enhancement}.

While recent work has advanced the quantification of melting rates and global heat transfer efficiency, comparatively little is known about the associated flow dynamics and hydrodynamic loading. In particular, detailed information on the time evolution of wake structures and force histories for melting bodies is scarce. Moreover, most numerical studies have focused on two-dimensional (2D) geometries such as circles and ellipses \citep{yang2024shape,yang2024circular}, leaving three-dimensional  (3D) effects largely unexplored. This gap is significant because the wake behind a non-melted rigid sphere undergoes well-documented bifurcations: a steady axisymmetric-to-planar transition near $\mathit{Re}\approx 212$ \citep{magnaudet2007wake}, a Hopf bifurcation to periodic vortex shedding near $\mathit{Re}\approx 273$ \citep{fabre2008bifurcations}, and ultimately a chaotic three-dimensional state at higher Reynolds numbers \citep{ern2012wake}. How these canonical wake transitions are altered when the sphere is melting and its shape evolves remains poorly understood.

Motivated by this question, we present three-dimensional direct numerical simulations of a melting sphere across the initial Reynolds number range $25 \leq \mathit{Re}_0 \leq 1000$. A sharp-interface method \citep{xue_three-dimensional_2023, xue2024flow}, based on a hybrid volume-of-fluid (VOF) and embedded-boundary (EB) method, is employed to resolve the flow and temperature fields in both liquid and solid phases separately, with interfacial conditions enforced explicitly. We consider a pure substance without solutal effects to focus on thermal melting under forced and mixed convection. The paper is organized as follows. The governing model and numerical approach are introduced in Section~\ref{sec:problem-stat}. Section~\ref{sec:melt-dynam-flow} examines flow patterns and interfacial morphologies across parameter regimes when buoyancy force is excluded. Section~\ref{sec:front-interface} quantifies melting rates, comparing with classical scaling laws and incorporating shape effects to improve prediction. Section~\ref{sec:drag-force-lift} analyses the time-dependent hydrodynamic forces and torques, while Section~\ref{sec:buoyancy} explores the influence of buoyancy-driven convection. Concluding remarks and perspectives are given in Section~\ref{sec:conclusion}.

\section{Problem statement, governing equations and numerical method}\label{sec:problem-stat}

\begin{figure}
  \centering
  \includegraphics[width=.8\textwidth]{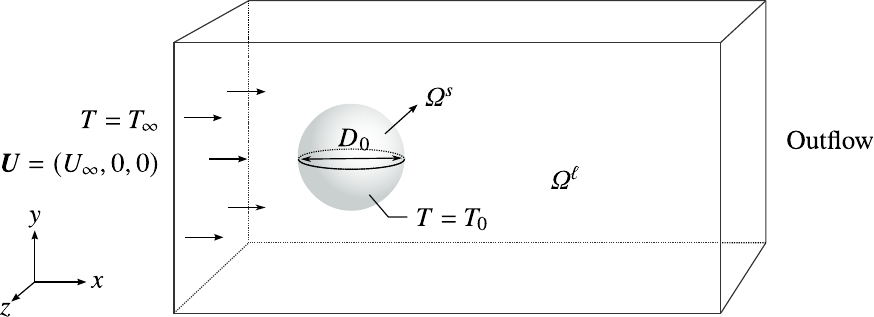}
  \caption{Schematic of the numerical setup (not to scale). A solid sphere composed of a pure substance with initial diameter $D_0$, denoted by $\Omega^s$, with uniform initial temperature $T = T_0$, is immersed in its warmer liquid phase, $\Omega^\ell$, driven by an incoming flow. At the inflow boundary, a uniform velocity $\bm{U} = (U_\infty, 0, 0)$ and temperature $T = T_\infty > T_0$ are imposed.}
  \label{fig:schema}
\end{figure}

The configuration considered is shown in figure~\ref{fig:schema}. An initially spherical solid particle ($\Omega^s$) of diameter $D_0$ and uniform temperature $T=T_0$ is fixed at the origin $\bm{x} = (x,y,z)= (0,0,0)$ and immersed in its warmer liquid phase ($\Omega^\ell$), representing the translational motion of a melting sphere. The liquid is incompressible and quiescent at $t=0$. A Cartesian coordinate system is adopted with the $x$-axis aligned with the streamwise direction. At the inflow boundary ($x=-4D_0$), uniform velocity $\bm{U} = (U_\infty, 0, 0)$ and temperature $T = T_\infty > T_0$ are prescribed, while an outflow condition is applied at $x=16D_0$. Lateral boundaries ($y=\pm 10D_0$, $z=\pm 10D_0$) are treated as adiabatic and free-slip.

Using $D_0$, $U_\infty$, and $\Delta T=T_\infty-T_0$ as characteristic length, velocity, and temperature scales, the governing equations for the liquid phase are

\begin{equation}\label{eq:ns1}
  \partial_t\bm{u} + \bm{u}\cdot\nabla\bm{u} = -\nabla p + \mathit{Re}_0^{-1}\nabla^2\bm{u} - \mathit{Ri}\,\theta\,\bm{e}_i, 
\end{equation} 
\begin{equation}\label{eq:ns2} 
\nabla\cdot\bm{u} = 0,
\end{equation}
\begin{equation}\label{eq:temperature}
  \partial_t\theta + \bm{u}\cdot\nabla\theta = (\mathit{Re}_0\mathit{Pr})^{-1} \nabla^2\theta,
\end{equation}
where $\bm{u}=(u_x,u_y,u_z)$ is the dimensionless velocity, $p$ the kinematic pressure, and $\theta=(T-T_0)/\Delta T$ the dimensionless temperature. The initial Reynolds, Prandtl, and Richardson numbers are defined as

\begin{equation}\label{eq:dimensionless_parameter}
\mathit{Re}_0 = \frac{U_\infty D_0 }{\nu}, \qquad \mathit{Pr} = \frac{\nu}{\alpha}, \qquad \mathit{Ri} = \frac{ (\bm{g}\cdot \bm{e}_i) \, \beta\,\Delta T\, D_0 }{U_\infty^2}, 
\end{equation} 
with $\nu$ the kinematic viscosity, $\alpha$ the thermal diffusivity, $\beta$ the thermal expansion coefficient, and $\bm{g}$ gravitational acceleration. Besides, the subscript `$0$' is adopted to distinguish it from a time-dependent effective Reynolds number $\mathit{Re}_e$ in the following analysis. For the Richardson number, no such distinction is needed since only the initial value appears in the analysis; hence we simply write $\mathit{Ri}$ without a subscript. \rec{The last term $-\mathit{Ri}\,\theta\,\bm{e}_i$ in (\ref{eq:ns1}) represents the thermal buoyancy force under the Boussinesq approximation. The unit vector $\bm{e}_i$ specifies the direction of gravity, either parallel ($i = x$) or perpendicular ($i = y$) to the streamwise direction. A linear dependence of liquid density on temperature is adopted here. The implications of using a more refined quadratic density–temperature relation are discussed in detail in \S~\ref{sec:buoyancy} and appendix \ref{sec:app}.} In the baseline simulations of the present study, buoyancy is neglected ($\mathit{Ri} = 0$), isolating forced convection that we mainly concern about. Mixed-convection cases, where buoyancy becomes comparable, are also examined in \S~\ref{sec:buoyancy}. Both phases are assumed to have identical constant density, so that volume expansion due to melting is ignored. All other thermophysical properties are constant in the liquid.

At the receding interface $\Gamma$, the no-slip condition $\bm{u}=\bm{0}$ is imposed, consistent with neglecting volume expansion during melting. The Gibbs–Thomson effect, which accounts for curvature- and velocity-dependent interface temperature, is omitted, and the interface temperature is fixed at the melting point, $\theta_\Gamma=0$, following the same principle of previous studies \citep{yang2024shape, yang2024circular, xue2024flow}. Consequently, the solid remains isothermal at $T^s \equiv T_0$ ($\theta^s\equiv0$) during melting. The interfacial motion is governed by the Stefan condition, which links the melting rate to the temperature gradient across the interface:
\begin{equation}\label{eq:interface-condition-2}
v_\mathit{\Gamma} = \mathit{St}(\mathit{Re}_0\mathit{Pr})^{-1}(\nabla\theta^\ell - \nabla\theta^s) \cdot \bm{n},
\end{equation}
where $v_\mathit{\Gamma}$ is the local interface velocity and $\bm{n}$ the outward unit normal pointing into the liquid. The Stefan number $\mathit{St} = c_p^\ell (T_\infty - T_0)/\mathcal{L}$ quantifies the ratio of sensible heat to latent heat, where $c_p^\ell$ is the specific heat capacity of the liquid and $\mathcal{L}$ is the latent heat of fusion. Since $\nabla\theta^s\equiv0$, (\ref{eq:interface-condition-2}) reduces to a condition in which the local melting rate depends solely on the temperature gradient on the liquid side.

\begin{figure}
  \centering
  \includegraphics[width=\textwidth]{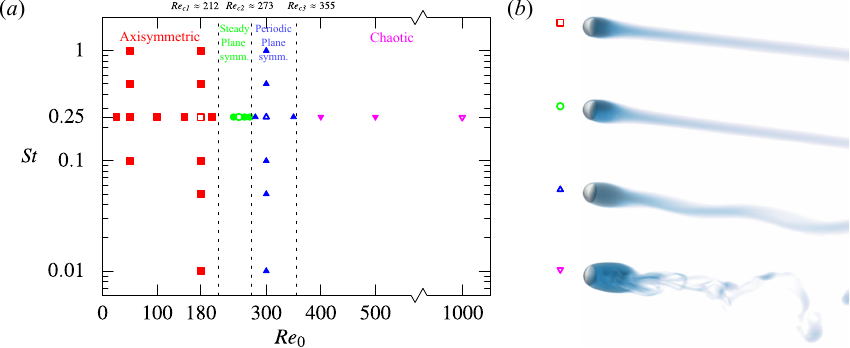}
  \caption{Parameter space $(\mathit{Re}_0, \mathit{St})$ explored in this study, with Prandtl number fixed at $Pr= 7$.
(\textit{a}) Range of initial Reynolds numbers, $25\leq \mathit{Re}_0\leq 1000$, covering the canonical regimes for non-melting spheres: steady axi-symmetric ($\mathit{Re}_0<212$), steady-planar-symmetric ($212<\mathit{Re}_0<273$), periodic-planar-symmetric ($273<\mathit{Re}_0<355$), and chaotic ($\mathit{Re}_0>355$), with regime boundaries from \citet{ern2012wake}. (\textit{b}) Representative snapshots of melting spheres and surrounding flow for each regimes, coloured by local temperature. The unfilled markers in the panel (\textit{a}) highlight corresponding visualisations.}
  \label{fig:regime}
\end{figure}

To illustrate the parameter space in physical terms, consider an ice sphere immersed in water at $T_\infty=20~^\circ\mathrm{C}$ and translating at $U_\infty=0.05~\mathrm{m/s}$. Varying its diameter from $0.5~\mathrm{mm}$ to $2~\mathrm{cm}$ yields $25\leq \mathit{Re}_0 \leq 1000$, spanning the above regimes shown in figure~\ref{fig:regime}(\textit{a}). The Prandtl number is fixed at $\mathit{Pr}=7$, representative of water at $20~^\circ\mathrm{C}$. Additionally, unless otherwise specified, the Stefan number is fixed at $\mathit{St}=0.25$, corresponding to $\Delta T = 20~^\circ\mathrm{C}$ in water, to isolate the effect of $\mathit{Re}_0$ on the melting dynamics. In \S\S~\ref{sec:front-interface} and \ref{sec:drag-force-lift}, $\mathit{St}$ is also varied in $[0.01,1]$ to quantify its influence on interface evolution and force history. Particularly, only the force convection is investigated in \S\S~\ref{sec:melt-dynam-flow}, \ref{sec:front-interface} and \ref{sec:drag-force-lift}, while mixed-convection effects are addressed separately in \S~\ref{sec:buoyancy}.

An important consideration concerns the initial flow development. Viscous and thermal boundary layers require finite time to establish, and wake structures do not appear immediately after flow initiation. Since the present study focuses on the interaction between melting and fully developed wakes, the melting process is initially deactivated and is only activated once the flow field reaches a steady or quasi-steady state, following the approach of \citet{yang2024shape}. This procedure avoids startup transients that might otherwise obscure the intrinsic coupling between convection and interface dynamics.

\rec{Due to the continuously evolving solid phase during melting, an important quantity is the time-dependent effective Reynolds number $\mathit{Re}_e(t)=\mathit{Re}_0 D_e(t)$, defined using an effective diameter $D_e(t)=2\sqrt{A_x(t)/\pi}$ calculated from the projected area in the projected cross-sectional area $A_x(t)$ in the incoming-flow ($x$--) direction. This parameter more clearly identifies the dynamical state of the system and facilitates the hydrodynamic analysis presented in \S~\ref{sec:drag-force-lift}.}

The governing equations~(\ref{eq:ns1})-(\ref{eq:temperature}), together with the Stefan condition~(\ref{eq:interface-condition-2}), are solved using a recently developed VOF-EB based sharp-interface method \citep{xue_three-dimensional_2023}. The scheme employs an EB framework that resolves the liquid and solid phases separately, with the interface being tracked and reconstructed using a VOF technique, and the interfacial jump conditions imposed sharply at $\Gamma$. The implementation is based on the open-source flow solver \textit{Basilisk} \citep{popinet2025basilisk}. Validation against laboratory measurements of melting ice spheres \citep{xue_three-dimensional_2023} demonstrated that the method reproduces both melting rates and interfacial evolution with high fidelity. 

\begin{figure}
  \centering
  \includegraphics[width=\textwidth]{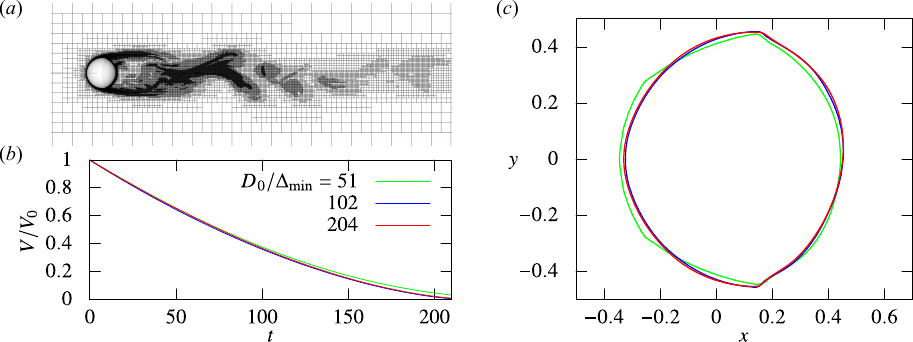}
  \caption{Grid refinement and numerical convergence for the case $(\mathit{Re}_0,\mathit{St},\mathit{Pr},\mathit{Ri})=(1000,0.25,7,0)$. (\textit{a}) Grid distribution on the mid-plane at $z=0$ at $t=30$ with $D_0/\Delta_\mathrm{min} = 204$, where $\Delta_\mathrm{min}$ denotes the finest mesh size.
(\textit{b}) Time evolution of normalized remaining solid volume $V(t)/V_0$ for varying spatial resolution $D_0/\Delta_\mathrm{min}$.
(\textit{c}) Solid-liquid interfaces on the mid-plane at $z=0$ at $t=60$ for varying spatial resolution $D_0/\Delta_\mathrm{min}$. The legend applies to both panels.}
  \label{fig:conv}
\end{figure}

Computational efficiency is achieved via adaptive mesh refinement, which concentrates resolution near the melting interface, where viscous and thermal boundary layers form, and in regions of strong vorticity in the wake. In the present simulations, the finest mesh size is $\Delta_{\min}=D_0/204$, ensuring at least seven grid points across both viscous and thermal boundary layers even at the highest initial Reynolds number ($\mathit{Re}_0=1000$). Figure~\ref{fig:conv}(\textit{a}) illustrates the resulting grid distribution around the sphere and in the wake. Grid convergence is assessed for the case $(\mathit{Re}_0,\mathit{St},\mathit{Pr},\mathit{Ri})=(1000,0.25,7,0)$. The normalized remaining solid volume $V(t)/V_0$ and solid-liquid interface at $t=40$ are compared across multiple resolutions in figures~\ref{fig:conv}(\textit{b},\textit{c}). The results show negligible difference between $\Delta_{\min}=D_0/204$ and $D_0/102$, confirming that the chosen resolution is sufficient to capture both interfacial and flow dynamics. 

\section{Structures of melting interface and fluid flow}\label{sec:melt-dynam-flow}

\subsection{$\mathit{Re}_0 < 212$: \textit{Axi-symmetric melting regime}}\label{sec:axi-symm-melt}

\begin{figure}
  \centering
  \includegraphics[width=\textwidth]{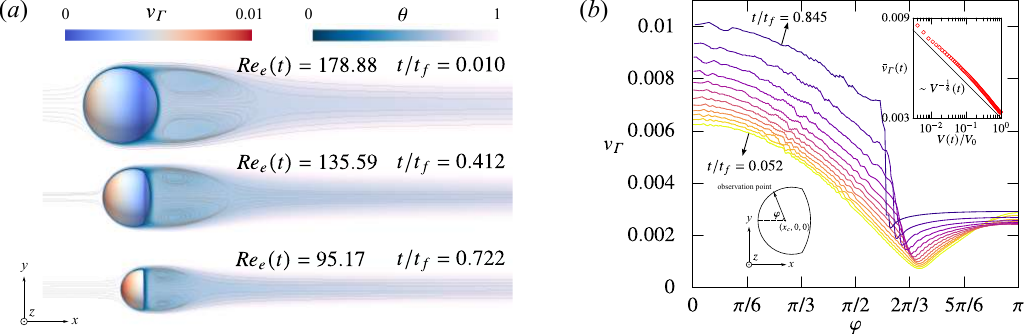}
  \caption{Melting dynamics at $\mathit{Re}_0=180$.
(\textit{a}) Snapshots of the temperature field $\theta$ and streamlines on the mid-plane at $z=0$, together with the 3D melting interface coloured by the local melting rate $v_\mathit{\Gamma}$, shown at $t/t_f = 0.010,0.412,0.722$ (top to bottom). \rec{The corresponding effective Reynolds numbers $\mathit{Re}_e(t)$ are indicated in each panel, and the same notation is used in the subsequent figures illustrating the melting process.}
\rec{(\textit{b}) Angular distribution of $v_\mathit{\Gamma}$ along the interface as a function of the polar angle $\varphi$, from $t/t_f=0.052$ to $0.845$ in increments of $\Delta t/t_f=0.072$. The inset at the lower left shows the polar angle $\varphi$; the origin is the body’s instantaneous mass centre $(x_c,0,0)$, and $\varphi$ is measured from the front stagnation point. The inset at the upper right shows the interface-averaged melting rate $\bar{v}_{\Gamma}(t)$ as a function of the normalised remaining volume $V(t)/V_0$, revealing a clear $-1/6$ scaling.
}
}
  \label{fig:axi-field}
\end{figure}

In the absence of melting, the wake behind a fixed sphere remains axi-symmetric up to the first bifurcation at $\mathit{Re}_{\mathit{c1}}\approx 212$ \citep{johnson_flow_1999, ern2012wake}. For a melting sphere with $\mathit{Re}_0<212$, both the interface and the surrounding flow are therefore expected to preserve axi-symmetry throughout the evolution. This is confirmed in figure~\ref{fig:axi-field}(\textit{a}) for $\mathit{Re}_0=180$, which displays the temperature field, mid-plane streamlines, and the 3D interface coloured by $v_\mathit{\Gamma}$. At all times shown, i.e. $t/t_f = 0.010$, $0.412$, and $0.722$ with $t_f$ denoting the moment at which the solid vanishes, the wake exhibits a symmetric recirculation zone and the interfacial shape remains axisymmetric. On the upstream side, $v_\mathit{\Gamma}$ decreases monotonically with the polar angle $\varphi$, as quantified in figure~\ref{fig:axi-field}(\textit{b}). This trend arises from boundary-layer development: starting from the stagnation point, the thermal boundary layer thickens with increasing arc length, thereby reducing the local heat flux. Since $v_\mathit{\Gamma} \propto \delta_\theta^{-1}$ (equation~\ref{eq:interface-condition-2}), the melting rate correspondingly decays with $\varphi$. By contrast, on the downstream side, flow separation isolates the rear surface from direct contact with the warmer inflow. The separated recirculation region traps relatively cold fluid, suppressing convective transport into the wake. As a result, the temperature gradient across the rear interface remains weak and $v_\mathit{\Gamma}$ exhibits an abrupt decline near the separation point, followed by persistently small values further downstream, as evident in figure~\ref{fig:axi-field}(\textit{b}).

As shown in figure~\ref{fig:axi-field}(\textit{a}), the front of the melting sphere approximately retains a spherical arc throughout the process, whereas the rear interface progressively flattens. This behaviour is consistent with previous experimental and numerical observations \citep{mac2015shape, yang2024shape, hao2001melting}, underscoring the close coupling between wake structure and rear-interface evolution. The flattening originates from the recirculating wake, which establishes a secondary stagnation point at the rear centre ($\varphi = \pi$). 
\rec{At this location, the recirculating wake produces a local stagnation point where the thermal boundary layer reaches its minimum thickness, yielding the peak melting rate $v_{\Gamma}$. This enhanced recession at $\varphi=\pi$ diminishes the rear curvature and drives the interface toward a planar shape. 
According to classical boundary-layer theory for axisymmetric stagnation-point flow \citep[][(5.70)]{schlichting_boundary-layer_2000}, a planar rear interface leads to an almost uniform viscous boundary-layer thickness. At $\mathit{Pr}=7$, the thermal layer lies within the viscous one and obeys the same scaling, so its thickness is likewise nearly uniform. This, in turn,
implies an approximately uniform melting rate along the entire rear surface.
Hence, once the interface becomes nearly flat, the boundary-layer structure naturally enforces the spatial uniformity of $v_{\Gamma}$ at the rear interface observed later in figure~\ref{fig:axi-field}(\textit{b}) at $t/t_f=0.845$, stabilising the planar configuration.}
This tendency mirrors the self-similar retreating shapes documented for dissolving or eroding bodies in high-Reynolds-number flows ($\mathit{Re}\sim10^4$), where nearly uniform wall shear or solute concentration profiles control the late-stage morphology \citep{ristroph2012sculpting, moore2013self, mac2015shape}. The present results thus suggest that similar boundary-layer mechanisms operate the rear interface in the moderate-$\mathit{Re}_0$ regime. Notably, as shown later in \S~\ref{sec:chao-melt-regime}, analogous uniformity also emerges at the front interface for high Reynolds number of $\mathit{Re}_0=1000$.

\rec{
In addition, figure~\ref{fig:axi-field}(\textit{b}) shows that the overall melting rate increases over time. This trend is further illustrated by the upper-right inset, which presents the interface-averaged melting rate as a function of the normalised remaining volume. The observed behaviour can be rationalised using classical boundary-layer scaling arguments \citep{schlichting_boundary-layer_2000,johnson_flow_1999,mac2015shape}. For a shrinking body, the viscous boundary-layer thickness satisfies the scaling $\delta_u / V^{1/3} \sim (V^{1/3})^{-1/2}$. Consequently, the interface-averaged melting rate scales as
\begin{equation}\label{eq:increasing-melting-rate-with-time}
\bar{v}_\mathit{\Gamma} \sim \delta_\theta^{-1} \sim \delta_u^{-1} \sim V^{-1/6},
\end{equation}
which agrees with the $-1/6$ scaling observed in the inset at the upper right of figure~\ref{fig:axi-field}(\textit{b}).
}

\begin{figure}
  \centering
  \includegraphics[width=\textwidth]{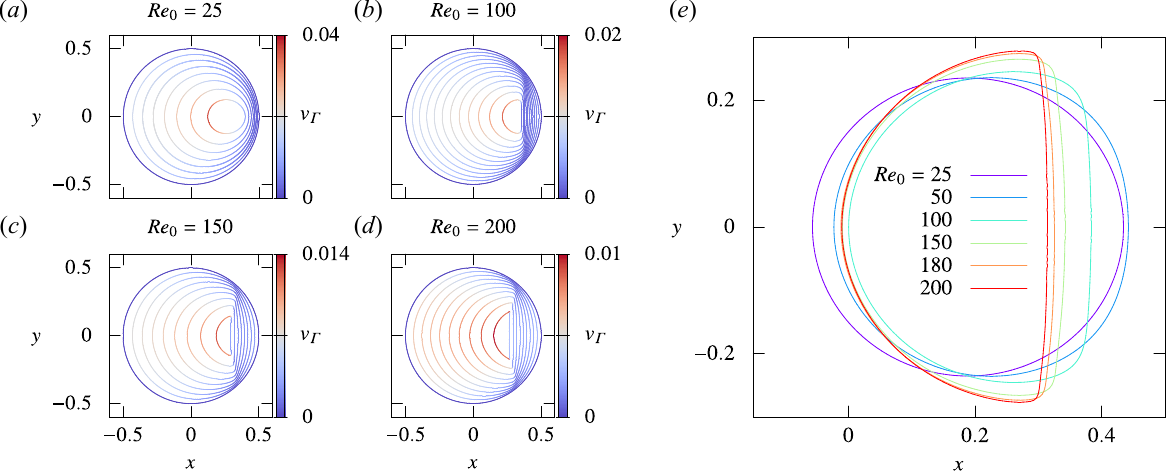}
  \caption{Effect of the initial Reynolds number $\mathit{Re}_0$ on interface evolution in the \textit{axi-symmetric melting regime}. (\textit{a–d}) Time evolution of the interface on the mid-plane at $z=0$ for $\mathit{Re}_0=25, 100, 150,$ and $200$, coloured by the local melting rate $v_\mathit{\Gamma}$. The sequences span nearly the entire melting process, with frames shown at uniform time intervals up to $t/t_f \approx 0.9$. (\textit{e}) Comparison of the interfaces on the mid-plane at $z=0$ when the remaining solid volume reaches $V(t)/V_0=0.1$, highlighting the progressive flattening of the rear surface as $\mathit{Re}_0$ increases.}
\label{fig:facet-comparison-V-0.1}
\end{figure}

The influence of the initial Reynolds number $\mathit{Re}_0$ on melting dynamics within the axi-symmetric regime is illustrated in figure~\ref{fig:facet-comparison-V-0.1}. For non-melting spheres, increasing $\mathit{Re}_0$ reduces the separation angle $\varphi_s$ and enlarges the toroidal wake vortex \citep{johnson_flow_1999}. A comparable trend emerges in the melting problem: larger $\mathit{Re}_0$ produces stronger wake recirculation, which progressively alters the rear-interface morphology. Figures~\ref{fig:facet-comparison-V-0.1}(\textit{a}-\textit{d}) show the mid-plane interface ($z=0$) at successive times for different $\mathit{Re}_0$, with the local melting rate $v_\mathit{\Gamma}$ indicated by colour. To enable direct comparison, figure~\ref{fig:facet-comparison-V-0.1}(\textit{e}) superposes the interfaces at the moment when the solid volume has decreased to $V(t)/V_0=0.1$. At moderate Reynolds numbers ($\mathit{Re}_0=25$ and $50$), the bodies preserve an approximately ellipsoidal form throughout melting, and the rear interface remains convex. In contrast, at higher $\mathit{Re}_0$ ($100 \lesssim \mathit{Re}_0 \leq 200$), the rear interface undergoes marked flattening, induced by stronger and more persistent recirculating wakes. This contrast stems from the reduced extent of separation at low-to-moderate $\mathit{Re}_0$, where the separation angle remains large (approximately $\varphi_s \approx 140^\circ$ for $\mathit{Re}_0 \leq 50$ \citep{johnson_flow_1999}), and from the decrease in effective Reynolds number $\mathit{Re}_e$ as the body shrinks. A more quantitative description of these shape changes is provided in \S~\ref{sec:front-interface}, where aspect ratio-based measures are introduced. The implications of interface morphology for global melting rates and hydrodynamic forces are analysed in \S\S~\ref{sec:front-interface} and \ref{sec:drag-force-lift}.

\subsection{$212 < \mathit{Re}_0 <  273$: \textit{Steady-planar-symmetric melting regime}}\label{sec:plane-symm-melt}

\begin{figure}
  \centering
  \includegraphics[width=.99\textwidth]{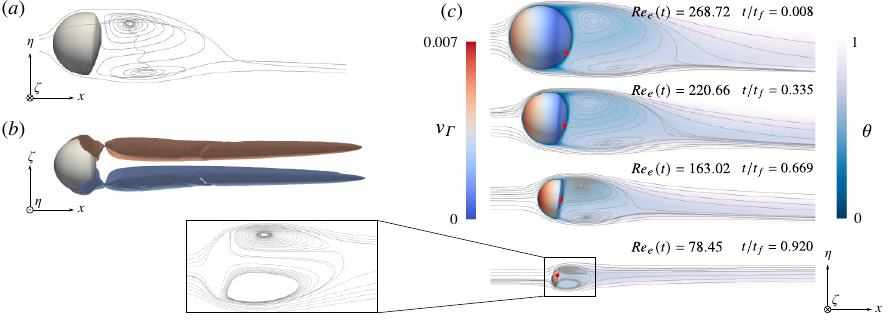}
  \caption{Melting dynamics at $\mathit{Re}_0=270$ in the \textit{steady-planar-symmetric regime}. 
  (\textit{a}) \reb{Three-dimensional streamlines viewed along the $\zeta$-direction.} (\textit{b}) \reb{Three-dimensional isosurfaces of the streamwise vorticity $\omega_x$ viewed along the $\eta$-direction, with red and blue surfaces indicating $\omega_x=\pm 0.25$.}
 \rec{The results shown in panels (\textit{a}) and (\textit{b}) correspond to $t/t_f = 0.502$, at which the effective Reynolds number is $\mathit{Re}_e(t)\approx 199.56$.} (\textit{c}) Snapshots of the temperature field $\theta$, streamlines on the symmetry plane ($x$--$\eta$), and the 3D interface colored by the local melting rate $v_\mathit{\Gamma}$, shown at $t/t_f=0.008$, $0.335$, $0.669$, and $0.920$. Red points mark the rear stagnation position, highlighting its lateral migration along $\eta$-direction and eventual reversal of spiral dominance in the wake.}
  \label{fig:plane-3D-melting-processes}
\end{figure}

For $\mathit{Re}_0 > 212$, the wake behind a translating sphere loses axi-symmetry but retains a reflectional symmetry plane, up to the second bifurcation threshold at $\mathit{Re}_{\mathit{c2}} \approx 273$ \citep{johnson_flow_1999, ern2012wake}. Within this steady-planar-symmetric regime, we examine the case $\mathit{Re}_0 = 270$. Since the flow spontaneously selects a symmetry plane, rather than being imposed, we analyze it using a rotated coordinate system $x$--$\eta$--$ \zeta$, where $x$--$\eta$ defines the symmetry plane and $\zeta$ is the transverse direction. \rea{Please note that, in the present study, the symmetry plane is entirely determined by numerical noise. As a result, the rotated coordinate system $x$–$\eta$–$\zeta$ used for the symmetry--breaking regimes throughout the manuscript is $\mathit{Re}_0$--dependent.} 
\reb{
Figure~\ref{fig:plane-3D-melting-processes}(\textit{a}) presents the three-dimensional streamlines in the vicinity of the sphere at $t/t_f = 0.502$, viewed along the $\zeta$-direction. The upstream fluid is first entrained by the lower spiral, then transported azimuthally toward the upper spiral, and finally guided around the lower spiral before moving downstream. Panel~(\textit{b}) shows the three-dimensional isosurfaces of the streamwise vorticity $\omega_x$ at the same instant, with red and blue surfaces representing $\omega_x = \pm 0.25$.
}
These illustrate the canonical features of the planar-symmetric wake: the recirculating vortex ring behind the sphere becomes tilted, with unequal upper and lower spirals, as illustrated in figure~\ref{fig:plane-3D-melting-processes}(\textit{c}) where the snapshots of the melting body and wake are presented. The stronger spiral entrains fluid into the weaker one, thereby breaking the closed toroidal bubble of the \textit{axi-symmetric melting regime}. This imbalance generates a pair of counter-rotating streamwise vortices, which induce a transverse lift force in the $\eta$-direction. Still in figure~\ref{fig:plane-3D-melting-processes}(\textit{c}), the front interface remains rounded, while the rear evolves asymmetrically: the lower side melts faster, producing an inclined planar surface that ultimately stabilizes at an angle of $-9.2^\circ$ relative to the $\eta$-direction. This asymmetry arises because the upper spiral traps colder recirculated fluid, suppressing local melting, whereas the lower spiral is replenished by warmer inflow, enhancing melting. The rear stagnation point (red markers) shifts laterally during the evolution, initially located at negative $\eta$ ($\eta < 0$ or $\varphi > \pi$) but crossing into positive $\eta$ ($\eta > 0$ or $\varphi < \pi$) at late times of $t/t_f = 0.920$, accompanied by a reversal of spiral dominance, as highlighted in the magnified view. This reversal can be attributed to the inclination of the melting body: the streamline path connecting the front stagnation point to the rear is shorter in the lower half, resulting in stronger vortex formation on that side. Such scenario is also consistent with vortex structures observed behind inclined non-melting bodies such as spheroids \citep{wang2021numerical} and cylinders \citep{kharrouba2021flow}. The implications of this spiral reversal for the lift force will be examined in \S~\ref{sec:lift-force}.

\begin{figure}
  \centering
  \includegraphics[width=.99\textwidth]{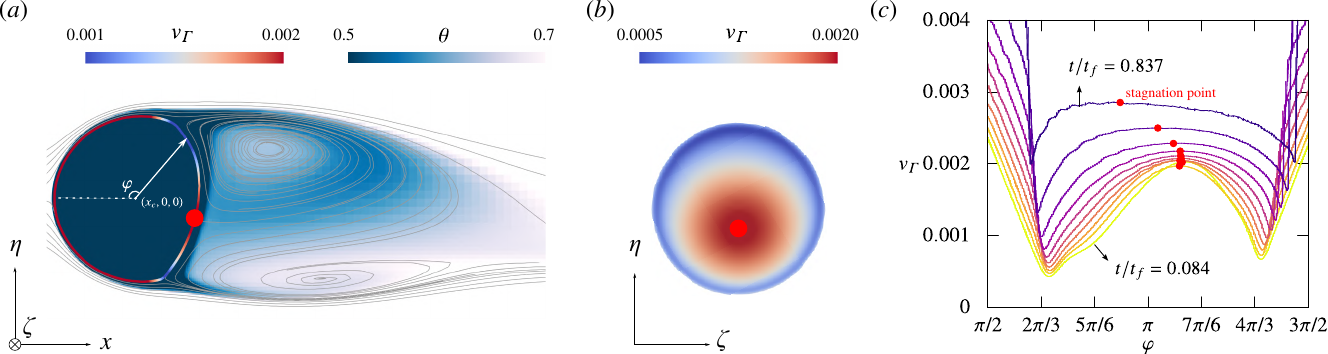}
  \caption{Rear-interface melting characteristics at $\mathit{Re}_0=270$. (\textit{a}) Temperature field $\theta$ and local melting rate $v_\mathit{\Gamma}$ on the symmetry plane ($x$--$\eta$). (\textit{b}) Distribution of $v_\mathit{\Gamma}$ viewed from the rear (positive $x$-axis). \rec{The results shown in panels (\textit{a}) and (\textit{b}) correspond to $t/t_f = 0.335$, at which the effective Reynolds number is $\mathit{Re}_e(t)\approx 220.66$.} (\textit{c}) Temporal evolution of $v_\mathit{\Gamma}$ as a function of the polar angle $\varphi$, from $t/t_f=0.084$ to $0.837$ at constant time intervals. Red circles mark the rear stagnation point in all panels.}
  \label{fig:plane-back-melting-rate}
\end{figure}

Figure~\ref{fig:plane-back-melting-rate} examines in detail the melting characteristics at $t/t_f=0.335$ for $\mathit{Re}_0=270$. Panel (\textit{a}) displays the temperature field $\theta$ and the local melting rate $v_\mathit{\Gamma}$ in the reflectional symmetry plane ($x$--$\eta$). A pronounced thermal asymmetry is observed: the lower spiral contains warmer fluid than the upper, consistent with the enhanced melting on the lower rear surface. The strongest melting occurs at the rear stagnation point, where the temperature gradient is steepest. This ``melting kernel” is highlighted in all panels by red circles. The rear-view distribution of $v_\mathit{\Gamma}$, shown in panel (\textit{b}), further emphasizes the asymmetry, with the lower hemisphere retreating more rapidly. Panel (\textit{c}) shows the temporal evolution of $v_\mathit{\Gamma}$ as a function of the polar angle $\varphi$ (defined in the inset of panel (\textit{a})), from $t/t_f=0.084$ to $0.837$ at intervals of $\Delta t/t_f=0.084$. The peak melting rate consistently coincides with the stagnation point, whose angular location gradually migrates toward larger $\eta$ with time, reflecting the evolving wake geometry we described in figure~\ref{fig:plane-3D-melting-processes}(\textit{b}). In addition, the spatial variation of $v_\mathit{\Gamma}$ across the rear interface becomes progressively smoother, reminiscent of the trend toward uniform boundary-layer-driven melting observed in the \textit{axi-symmetric melting regime}. This suggests that, even in the presence of reflectional asymmetry, the system dynamically reorganizes to reduce rear-interface thermal gradients.

\begin{figure}
  \centering
  \includegraphics[width=.99\textwidth]{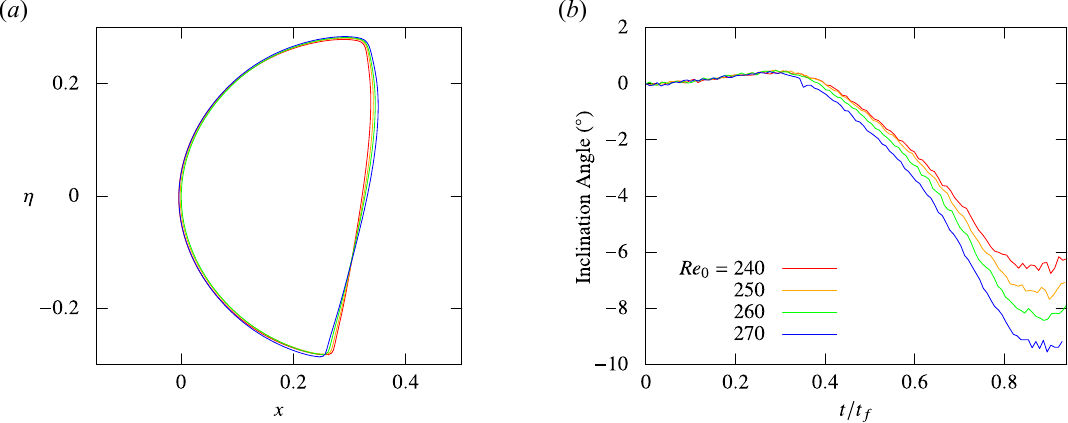}
  \caption{Influence of $\mathit{Re}_0$ on rear-interface evolution in the \textit{steady-planar-symmetric melting regime}. (\textit{a}) Comparison of interfaces on the symmetry plane ($x$--$\eta$)  when the remaining solid volume reaches $V(t)/V_0=0.1$. (\textit{b}) Time evolution of the inclination angle of the rear interface relative to the $x$-axis for different $\mathit{Re}_0$. Increasing $\mathit{Re}_0$ enhances the inclination of the rear interface.}
  \label{fig:plane-back-planar-interface}
\end{figure}

Figure~\ref{fig:plane-back-planar-interface} summarizes the influence of the initial Reynolds number $\mathit{Re}_0$ on the morphological evolution of the rear interface within the \textit{steady-planar-symmetric melting regime}, considering cases with $\mathit{Re}_0 = 240$, $250$, $260$, and $270$. As $\mathit{Re}_0$ increases, the wake asymmetry intensifies, producing progressively larger inclination angles of the rear interface relative to the $x$-direction. This trend is illustrated in panel (\textit{a}), which overlays the interfaces at a fixed volume fraction $V(t)/V_0=0.1$. While the front hemispheres collapse onto nearly identical spherical arcs, the diverging rear interfaces clearly demonstrate the increasing anisotropy of wake-induced melting at higher Reynolds numbers. To quantify this inclination, we define the rear-interface angle as $\mathrm{tan}^{-1}(\mathrm{sgn}(\bar{n}_\eta)\sqrt{\bar{n}^2_\eta+\bar{n}^2_\zeta}/|\bar{n}_x|)$, where $(\bar{n}_x,\bar{n}_\eta,\bar{n}_\zeta)$ are the components of the area-averaged outward normal vector $\bar{\bm{n}}$ on the rear interface. The time evolution of the rear-interface angle, shown in panel (\textit{b}), confirms this trend: the inclination increases from about $-6.2^\circ$ at $\mathit{Re}_0=240$ to $-9.2^\circ$ at $\mathit{Re}_0=270$. At late stages ($t/t_f > 0.8$), the inclination saturates, consistent with the nearly uniform rear melting rates observed earlier (see figure~\ref{fig:plane-back-melting-rate}(\textit{c})).

\subsection{$273 < \mathit{Re}_0 < 355$: \textit{Periodic-planar-symmetric melting regime}}\label{sec:peri-melt-regime}

\begin{figure}
  \centering
  \includegraphics[width=\textwidth]{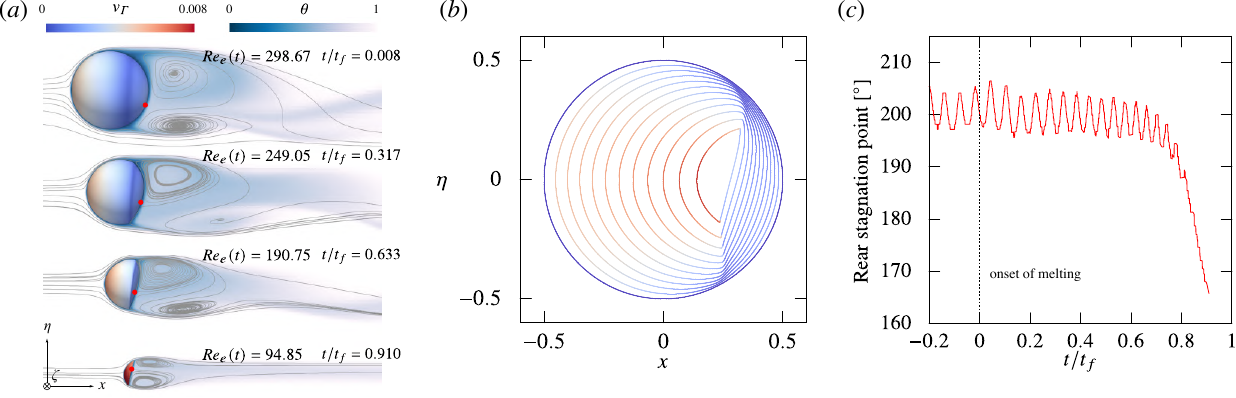}
  \caption{Melting processes of $\mathit{Re}_0=300$ in \textit{periodic-planar-symmetric melting regime}. (\textit{a}) Snapshots of the temperature field $\theta$ and streamline at the symmetry plane ($x$--$\eta$) and 3D interface coloured by the melting rate $v_\mathit{\Gamma}$ at $t=1,40,80,115$ ($t/t_f=0.008,0.317,0.633,0.910$), from top to bottom. The red points indicate the rear stagnation points at each moment. (\textit{b}) Time evolution of the interface colored by the melting rate $v_\mathit{\Gamma}$ at the reflectional symmetry plane ($x$--$\eta$) from $t=0$ to $t=110$ ($t/t_f=0.871$) with a constant time interval of $\Delta t=10$ ($\Delta t/t_f=0.079$). (\textit{c}) Time evolution of polar angle of the rear stagnation point, where the dynamical suppression of wake periodicity is observed as the effective Reynolds number $\mathit{Re}_e(t)$ decreases.}
  \label{fig:periodic-melting-processes}
\end{figure}

When the Reynolds number surpasses the second critical threshold, $\mathit{Re}_{\mathit{c2}} \approx 273$, the wake undergoes a Hopf bifurcation, transitioning from steady to unsteady behavior. The resulting flow exhibits time-periodic vortex shedding while retaining planar reflectional symmetry \citep{johnson_flow_1999}. In this regime, the wake generates oscillatory lift forces with a non-zero mean, yet the vortex separation line remains nearly fixed and the rear stagnation point fluctuates only weakly within $\pm 9^\circ$ \citep{johnson_flow_1999}. Thus, despite the periodic shedding, the rear stagnation point is effectively quasi-stationary.

Accordingly, we anticipate that once melting commences, the rear interface will continue to evolve into a planar and inclined surface, in close analogy with the behavior observed in the  \textit{steady-planar-symmetric melting regime} (\S~\ref{sec:plane-symm-melt}). This expectation is confirmed in figure~\ref{fig:periodic-melting-processes}(\textit{a}), which presents representative snapshots of the streamline topology and temperature field $\theta$ on the symmetry plane ($x$--$\eta$) for $\mathit{Re}_0 = 300$, together with the three-dimensional interface colored by the local melting rate $v_\mathit{\Gamma}$, over the interval $0 < t/t_f < 0.910$. Once melting is activated, the rear interface begins to incline, and the wake gradually loses its temporal periodicity, reorganizing into a new steady asymmetric configuration. By $t/t_f = 0.910$, the lower spiral has become larger than the upper, a reversal analogous to that observed in the \textit{steady-planar-symmetric melting regime} (\S~\ref{sec:plane-symm-melt}). The temporal development of this melting-induced transition is further illustrated in figure~\ref{fig:periodic-melting-processes}(\textit{b}), which shows that a tilted and nearly planar rear interface persists in the later stage. The dynamical suppression of wake periodicity is quantified in figure~\ref{fig:periodic-melting-processes}(\textit{c}), where the time histories of the polar angle of the rear stagnation point are plotted. \reb{These quantities} initially exhibit small oscillations, consistent with the underlying unsteady wake. As melting progresses, the effective Reynolds number $\mathit{Re}_e(t)$ decreases, leading to a decay of these oscillations and a monotonic shift toward smaller $\varphi$ values, \reb{corresponding} to the positive $\eta$-direction. This behavior demonstrates that interfacial evolution suppresses vortex shedding and re-establishes a quasi-steady wake morphology.

\begin{figure}
  \centering
  \includegraphics[width=\textwidth]{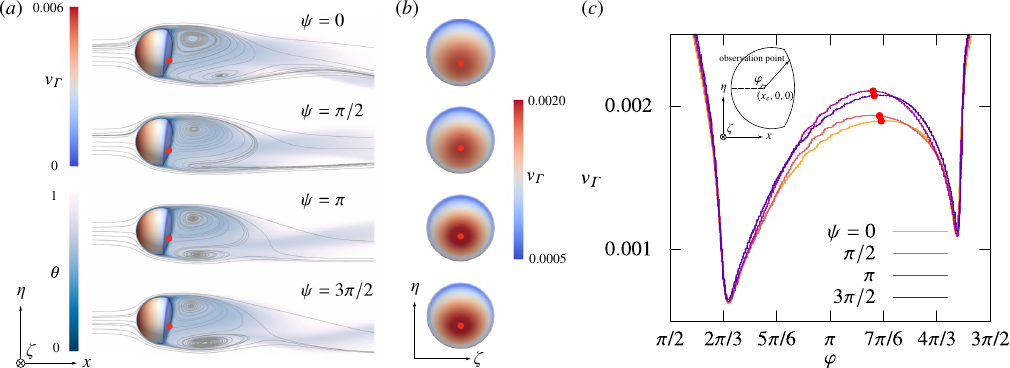}
  \caption{Phase-resolved melting dynamics during one oscillation cycle at $\mathit{Re}_0=300$. Four phases are shown, $\psi = 0, \pi/2, \pi, 3\pi/2$, corresponding to $t = 64.1, 65.5, 66.9, 68.3$ ($t/t_f=0.507, 0.519, 0.530, 0.540$). \rec{During this interval, the effective Reynolds number $\mathit{Re}_e(t)$ decreases from about $194.36$ to $189.02$.} (\textit{a}) Temperature field $\theta$ and streamlines on the symmetry plane ($x$--$\eta$), together with the 3D interface colored by the local melting rate $v_\mathit{\Gamma}$. (\textit{b}) Distribution of $v_\mathit{\Gamma}$ on the rear face of the solid, viewed from the rear. (\textit{c}) Polar distribution of $v_\mathit{\Gamma}$ at the symmetry plane ($x$--$\eta$)  for the four phases, where the inset defines the polar angle $\varphi$ for this regime. Red points denote the rear stagnation point in all panels.}
  \label{fig:periodic-one-period}
\end{figure}

Figure~\ref{fig:periodic-one-period} presents a phase-resolved view of the melting dynamics during a single oscillation cycle in the early stage of melting. Four characteristic phases are selected within one shedding period, $\psi = 0$, $\pi/2$, $\pi$, and $3\pi/2$, corresponding to $t = 64.1$, $65.5$, $66.9$, and $68.3$ ($t/t_f = 0.507, 0.519, 0.530, 0.540$). Panel~(\textit{a}) shows the temperature field and streamlines on the symmetry plane ($x$--$\eta$), together with the interface coloured by $v_\mathit{\Gamma}$. While the global wake remains quasi-steady, small periodic modulations are evident, reflecting the influence of vortex shedding on local convective transport. Panel~(\textit{b}) illustrates the rear-face distribution of $v_\mathit{\Gamma}$, which exhibits clear phase-dependent variations: the peak melting rate oscillates in sync with the shedding cycle. This behaviour arises from periodic changes in the thickness of the thermal boundary layer at the rear stagnation zone, which governs local heat flux. Panel~(\textit{c}) confirms this interpretation by showing the polar distribution of $v_\mathit{\Gamma}$ on the symmetry plane ($x$--$\eta$). At $\psi = 0$ ($t=64.1$), the lower spiral is in the process of detachment, which weakens convective transport near the rear surface and reduces the melting rate. In contrast, at $\psi = \pi$ ($t=66.9$), the lower spiral is replenished by inflow and remains attached to the surface, intensifying local convection and enhancing the melting rate.

\subsection{$\mathit{Re}_0 > 355$: \textit{Chaotic melting regime}}\label{sec:chao-melt-regime}

When the initial Reynolds number $\mathit{Re}_0$ exceeds the third critical threshold, $\mathit{Re}_{\mathit{c3}} \approx 355$, the wake behind a non-melting sphere transitions into a fully three-dimensional chaotic state \citep{mittal_planar_1999}. For a melting sphere in this regime, the rear stagnation point, typically associated with the maximum local melting rate, hence no longer remains spatially or temporally fixed, but undergoes irregular fluctuations. Consequently, the rear interface cannot evolve into a stable planar surface, in contrast with the behaviors observed in the \textit{steady-} and \textit{periodic-planar-symmetric melting regimes}.

\begin{figure}
  \centering
  \includegraphics[width=.95\textwidth]{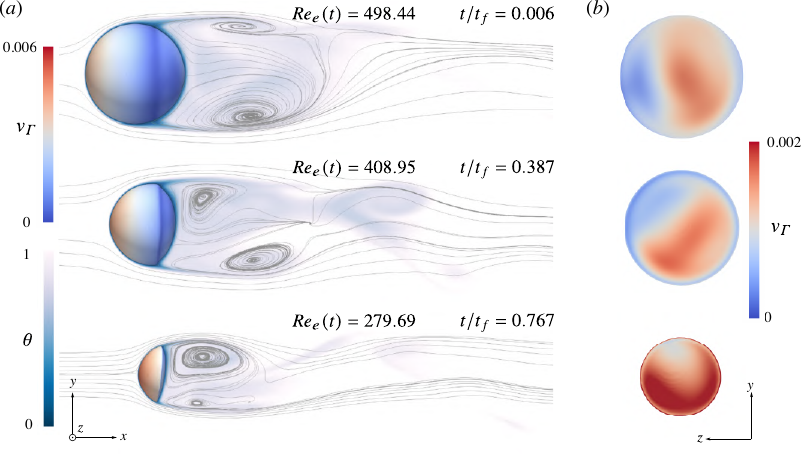}
  \caption{Melting dynamics for $\mathit{Re}_0 = 500$. (\textit{a}) Snapshots of the temperature field $\theta$ and streamlines on the mid-plane at $z=0$, together with the 3D interface colored by the local melting rate $v_\mathit{\Gamma}$, at $t=1,62,123$ ($t/t_f=0.006,0.387,0.767$). (\textit{b}) Corresponding melting rate distributions on the rear surface.}
  \label{fig:chaotic-melting-processes-re-500}
\end{figure}

\begin{figure}
  \centering
  \includegraphics[width=.95\textwidth]{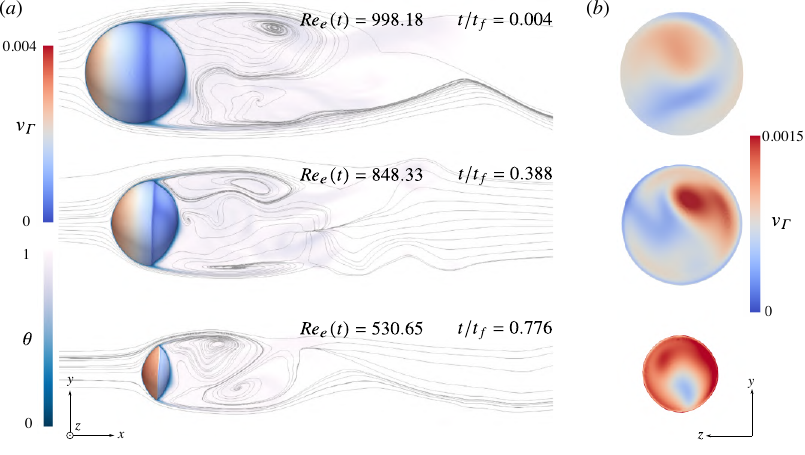}
  \caption{As in figure~\ref{fig:chaotic-melting-processes-re-500}, but for $\mathit{Re}_0=1000$. Snapshots are shown at $t=1,90,180$ ($t/t_f=0.004,0.388,0.776$).}
  \label{fig:chaotic-melting-processes-re-1000}
\end{figure}

Figures~\ref{fig:chaotic-melting-processes-re-500} and \ref{fig:chaotic-melting-processes-re-1000} confirm these expectations. For both $\mathit{Re}_0=500$ and $1000$, panel (\textit{a}) show mid-plane temperature and streamline fields superimposed on the 3D interface coloured by $v_\mathit{\Gamma}$, while panel (\textit{b}) display the corresponding rear-face melting rate distributions. Here we adopt the Cartesian $x$--$y$--$z$ coordinates rather than $x$--$\eta$--$\zeta$, since no clear planar symmetry persists. Unlike at $\mathit{Re}_0=300$ (figure~\ref{fig:periodic-melting-processes}), the wakes now exhibit strongly disordered vortex shedding and intense small-scale turbulence, characteristic of chaotic flow. At $\mathit{Re}_0=500$, the rear surface remains rounded only up to $t/t_f \leq 0.387$, whereas at $\mathit{Re}_0=1000$ a rounded rear persists throughout. This difference reflects the role of small-scale vortices generated by Kelvin–Helmholtz-type instabilities at boundary-layer separation \citep{tomboulides2000numerical}. These vortices disrupt large-scale coherence, suppress organized recirculation, and render the rear stagnation point highly unsteady in both space and time. Consequently, the peak in $v_\mathit{\Gamma}$ is spatially diffuse, and the interface never reorganizes into a planar rear—sharply contrasting the lower-$\mathit{Re}_0$ regimes (\S~\ref{sec:axi-symm-melt}–\S~\ref{sec:peri-melt-regime}), where a relatively stable stagnation point drives the formation of a well-defined planar interface.

An exception to this chaotic pattern is observed at $t/t_f = 0.767$ for the case $\mathit{Re}_0 = 500$ (figure~\ref{fig:chaotic-melting-processes-re-500}(\textit{a})), where a weakly inclined rear interface emerges, oriented towards the negative $y$-axis. This phenomenon can be attributed to the progressive reduction of the effective Reynolds number $\mathit{Re}_e(t)$. As $\mathit{Re}_e(t)$ falls below the chaotic threshold, the wake transitions back into a periodic, and eventually steady regime, thereby restoring partial symmetry and enabling planar interface development. Despite the irregularity of the wake in the \textit{chaotic regime}, figures~\ref{fig:chaotic-melting-processes-re-500}(\textit{b}) and \ref{fig:chaotic-melting-processes-re-1000}(\textit{b}) reveal a gradual increase in the spatial uniformity of $v_\mathit{\Gamma}$ across the rear surface as melting progresses. This observation reinforces the general principle, established at lower Reynolds numbers, that the system dynamically reorganizes to reduce gradients in the melting rate - a trend that persists even in high-$\mathit{Re}_0$ turbulent environments.

\begin{figure}
    \centering
    \includegraphics[width=0.9\textwidth]{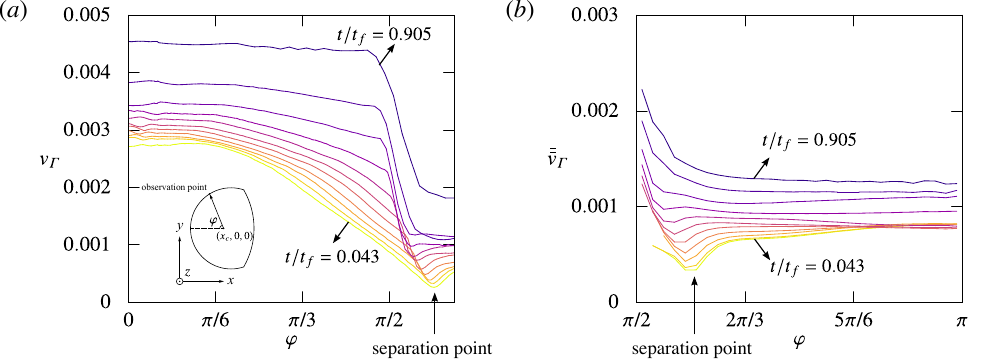}
    \caption{Local melting rate distribution for the case $\mathit{Re}_0=1000$ from $t=10$ ($t/t_f=0.043$) to $t=210$ ($t/t_f=0.905$) with a constant time interval $\Delta t=20$ ($\Delta t/t_f=0.086$). (\textit{a}) $v_\mathit{\Gamma}$ as a function of the polar angle $\varphi$ defined by the inset at the front on mid-plane at $z=0$. (\textit{b}) Tempo-spatially averaged melting rate $\bar{\bar{v}}_\mathit{\Gamma}$ on the rear interface as a function of the polar angle $\varphi$. Details of the averaging procedure are given in the text.}
    \label{fig:chaotic-mr-profile}
\end{figure}

Figure~\ref{fig:chaotic-mr-profile} provides a quantitative view of this tendency for $\mathit{Re}_0=1000$, with panels (\textit{a}) and (\textit{b}) showcasing the evolution of the melting rate along the front ($0<\varphi<\pi/2$) and the back ($\pi/2<\varphi<\pi$) interfaces, respectively. $\bar{\bar{v}}_\mathit{\Gamma}(t)$ is evaluated by averaging the local $v_\mathit{\Gamma}(t)$ over a narrow angular band ($\varphi \pm 1.5^\circ$) and a time window ($t \pm 10$). We also remind that the results are nearly identical for the opposite side ($-\pi<\varphi<0$) due to the uniformity of the melting in this chaotic flow regime, and hence they are not shown here. The resulting distributions in both pictures confirm that, even amid turbulence, the interface evolution systematically promotes homogenization of $v_\mathit{\Gamma}$ over time. More specifically, the chaotic wake dynamics do not preclude self-organization of melting: instead, the melting rate evolves toward greater spatial uniformity, with homogenization extending from the rear to the front interface at $\mathit{Re}_0=1000$ - a feature absent at lower Reynolds numbers ($\mathit{Re}_0 \leq 300$) but in consistent with the dissolution process at high $Re_0$ regime reported by \cite{mac2015shape}. Besides confirming that a melting body tends to converge toward a shape with nearly uniform material removal along its surface, it is further demonstrated \citep{moore2013self, moore2017riemann} that this principle inherently promotes the formation of a smooth, rounded front interface - a prediction that is fully consistent with our numerical observations.

\reb{At the end of this section, a natural question that arises is whether the wake bifurcations observed during melting can still be related to the classical critical Reynolds numbers established for non-melting rigid spheres, namely $\mathit{Re}_{c1}$, $\mathit{Re}_{c2}$ and $\mathit{Re}_{c3}$ in figure \ref{fig:regime}(\textit{a}). Our results show that this correspondence no longer holds. For instance, the $\mathit{Re}_0=300$ case loses vortex shedding only at $\mathit{Re}_e \approx 138$ ($\mathit{Re}_{c2}\approx 273$), the $\mathit{Re}_0=500$ case leaves the chaotic regime around $\mathit{Re}_e \approx 206$ ($\mathit{Re}_{c3}\approx 355$), and the $\mathit{Re}_0=1000$ case retains chaotic features throughout the melting process and only begins to develop a weakly time-periodic behaviour near $\mathit{Re}_e \approx 268$.
This systematic reduction of critical Reynolds numbers can be traced to two coupled mechanisms. First, the wake exhibits a hysteresis-like response: in a small test where the melting process was artificially stopped in the $Re_0=300$ case, vortex shedding persisted for several additional periods even though the instantaneous $Re_e$ had fallen well below the corresponding rigid-sphere Hopf bifurcation threshold, demonstrating a delayed decay of unsteady modes. Second, as the solid melts, its shape becomes increasingly disk-like, and such an increase in aspect ratio is known to reduce the bifurcation Reynolds numbers \citep{ern2012wake}. These combined effects explain why the classical critical Reynolds numbers cannot be directly applied to melting spheres.
}

\section{Shape dynamics and scaling laws}\label{sec:front-interface}

A quantitative description of the melting dynamics requires a predictive model for the time evolution of the remaining solid volume. In general, the volumetric decay rate can be expressed as
\begin{equation}\label{eq:volume-equation}
\frac{\mathrm{d}V(t)}{\mathrm{d}t} \sim \bar{v}_\mathit{\Gamma}(t) A(t),
\end{equation}
where $V(t)$ is the remaining solid volume, $A(t)$ is the instantaneous surface area of the interface, and $\bar{v}_\mathit{\Gamma}(t)$ denotes the spatially averaged interface melting rate. Closing (\ref{eq:volume-equation}) requires scaling relations for three key quantities: the total melting time $t_f$, the temporal evolution of $\bar{v}_\mathit{\Gamma}$, and the variation of $A(t)$. 

\begin{figure}
  \centering
  \includegraphics[width=.9\textwidth]{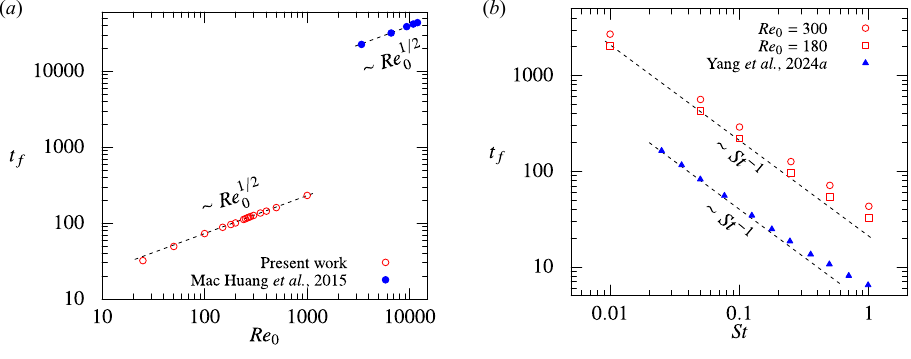}
  \caption{\reb{Dependence of the complete melting time $t_f$ on the initial Reynolds number $\mathit{Re}_0$ and the Stefan number $\mathit{St}$. (\textit{a}) Variation of $t_f$ as a function of $\mathit{Re}_0$ at $\mathit{St}=0.25$, where the experimental results from \cite{mac2015shape} are also shown by considering $D_0=6~\mathrm{cm}$ and $\nu=2\times 10^{-6}~\mathrm{m^2/s}$ in their dissolving processes. (\textit{b}) Variation of $t_f$ as a function of $\mathit{St}$ for $\mathit{Re}_0=180$ and $300$, where we also present the 2D numerical results of melting cylinders from \cite{yang2024shape}.}}
  \label{fig:tf-re-st}
\end{figure}

We first consider the scaling of the complete melting time $t_f$. 
\rec{
Note that the total melting time $t_f$ for each case is not obtained directly from the numerical simulation. As the solid shrinks, the three-dimensional solid–liquid interface is reconstructed from piecewise VOF facets, and the accuracy of this reconstruction degenerates when the remaining solid volume becomes very small. This degradation affects both the interpolation and the enforcement of jump conditions at the interface.
To circumvent this issue, we estimate $t_f$ by extrapolating from the moment when $1\%$ of the solid remains, i.e. when $V(t_{99})/V_0 = 0.01$, at which the 3D interface representation is still sufficiently accurate. The final melting time is then inferred from the relation $t_f = t_{99}/0.9$, which follows from the classical quadratic scaling (\ref{eq:classical-scaling-law-prediction}), namely $V(t)/V_0 = (1 - t/t_f)^2$, as will be established shortly.
}
Back to the scaling of $t_f$.
From the Stefan condition (\ref{eq:interface-condition-2}), the characteristic propagation speed of the interface scales as $\bar{v}_\mathit{\Gamma} \sim \mathit{St}\mathit{Re}_0^{-1}\delta_\theta^{-1}$. Boundary-layer theory predicts $\delta_\theta \sim \delta_u \sim \mathit{Re}_0^{-1/2}$ \citep{schlichting_boundary-layer_2000}. Combining these relations, the inverse dependence of $\bar{v}_\mathit{\Gamma}$ on melting time cancels partially, yielding the scaling law
\begin{equation}\label{eq:tf-scaling-law}
  t_f \sim \mathit{Re}_0^{1/2}\mathit{St}^{-1}.
\end{equation}

The prediction (\ref{eq:tf-scaling-law}) is confirmed numerically in figure~\ref{fig:tf-re-st}: panel (\textit{a}) shows $t_f \propto \mathit{Re}_0^{1/2}$ at fixed $\mathit{St}$, while panel (\textit{b}) demonstrates $t_f \propto \mathit{St}^{-1}$ at fixed $\mathit{Re}_0$. \reb{The scalings with respect to $\mathit{Re}_0$ and $\mathit{St}$ have also been proposed and verified in the experimental study of  dissolving bodies in fluid flow \citep{mac2015shape} and in the numerical work of melting cylinders under forced convection \citep{yang2024shape}, respectively. Additionally, we include their results in figure~\ref{fig:tf-re-st}. The agreement validates (\ref{eq:tf-scaling-law}) within a large parameter space.}

\red{
The scaling $t_f \sim \mathit{Re}_0^{1/2}$ may appear counter-intuitive, since one might expect the physical melting time to decrease as the incoming flow strength increases. This apparent contradiction arises because the dimensionless time $t_f$ is normalised by the advective time scale $D_0 / U_\infty$. 
Alternatively, one may normalise the melting time using the constant thermal diffusive time scale $D_0^2 / \alpha$, which more directly reflects the accelerated melting under stronger flows. The corresponding physical melting time $t_{f,\alpha}$ becomes
}

\begin{equation}
    \red{t_{f,\alpha} = \frac{t_f D_0/U_\infty}{D_0^2/\alpha} = t_f \frac{\nu}{D_0 U_\infty} \frac{\alpha}{\nu} \sim t_f \mathit{Re}_0^{-1} \sim \mathit{Re}_0^{-1/2} \mathit{St}^{-1}},
\end{equation}
\red{
which clearly indicates a negative correlation between the melting time and the incoming flow velocity.
This $-1/2$ scaling is also proposed by \cite{mac2015shape}, who reported a similar inverse dependence between the experimentally measured physical melting time and the flow velocity.
}

\begin{figure}
  \centering
  \includegraphics[width=\textwidth]{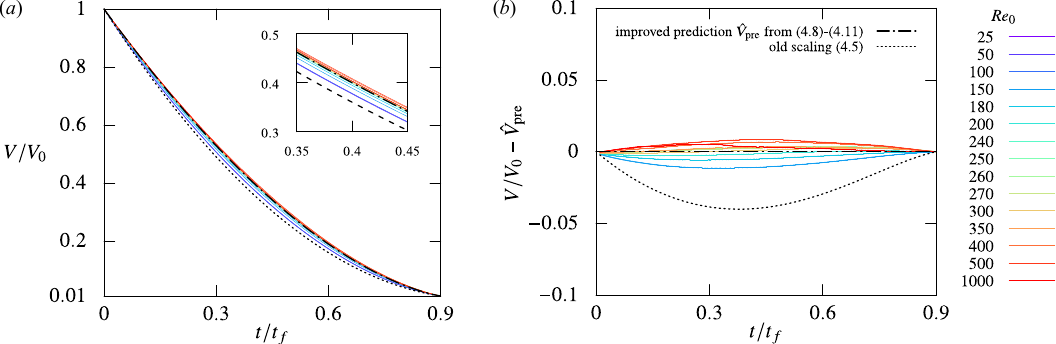}
  \caption{Time evolution of the normalized remaining solid volume $V/V_0$ for various $\mathit{Re}_0$. The data are compared against the old scaling $V/V_0 = (1-t/t_f)^2$ (dotted line) obtained from (\ref{eq:classical-scaling-law-prediction}) and the improved prediction $\hat{V}_{\mathrm{pre}}(\hat{t})$ (dash-dotted line) obtained from (\ref{eq:further-pred-init}) - (\ref{eq:further-pred-final}), respectively. (\textit{b}) Deviations of $V/V_0$ from the improved prediction for $\mathit{Re}_0 \geq 150$, with the old scaling shown for reference.}
  \label{fig:volume}
\end{figure}

\rec{
Assuming the total surface area scales as $A(t) \sim V^{2/3}(t)$ and integrating the correlation (\ref{eq:increasing-melting-rate-with-time}), the volumetric decay law (\ref{eq:volume-equation}) forms an ordinary differential equation (ODE):
\begin{equation}\label{eq:ode-for-volume}
  \frac{\mathrm{d}V}{\mathrm{d}t} \sim  \bar{v}_\mathit{\Gamma}(t) A(t)
  \sim V^{-1/6}(t) V^{2/3}(t) \sim V^{1/2}(t),
\end{equation}
whose integration yields the classical result
\begin{equation}\label{eq:classical-scaling-law-prediction}
  \frac{V(t)}{V_0} = (1-\frac{t}{t_f})^2,
\end{equation}
with global melting rate $\bar{v}_\Gamma \sim (1-t/t_f)^{-1/3}$.}
This quadratic decay law has been widely reported for eroding and melting bodies \citep{mac2015shape, yang2024shape}, and figure~\ref{fig:volume}(\textit{a}) shows that it reproduces our results well at moderate Reynolds numbers ($\mathit{Re}_0 < 100$). However, at larger Reynolds numbers ($\mathit{Re}_0 > 100$), systematic deviations appear: the quadratic law over-predicts the rate of volume loss, as evident in figure~\ref{fig:volume}(\textit{b}), which plots the deviation of the numerical data from the prediction (\ref{eq:classical-scaling-law-prediction}). 

\begin{figure}
  \centering
  \includegraphics[width=\textwidth]{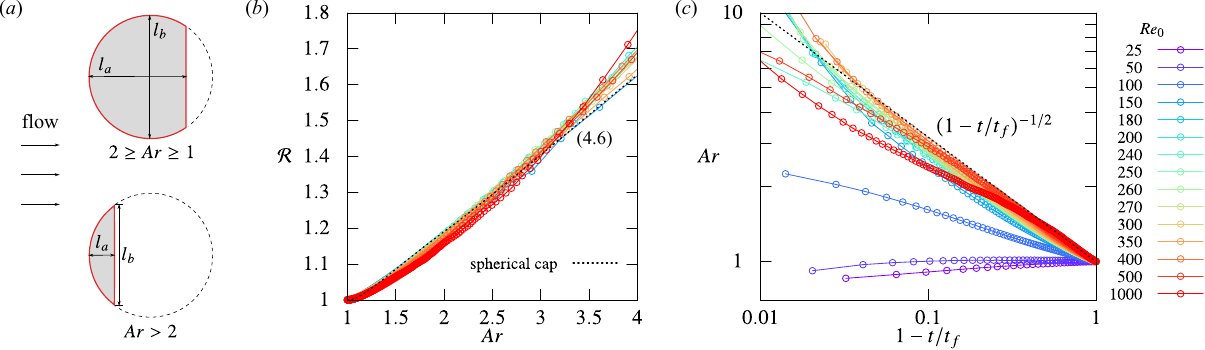}
  \caption{(\textit{a}) \rec{Spherical-cap model used to approximate the melting solid geometry for $\mathit{Re}_0 \geq 150$. Upper panel: shapes with $1 \leq \mathit{Ar} \leq 2$; lower panel: elongated shapes with $\mathit{Ar} > 2$. The aspect ratio is defined as $\mathit{Ar} = l_b/l_a$, where $l_a$ and $l_b$ are the streamwise and transverse extents. Note that this figure presents a 2D schematic of the 3D solid shape.} (\textit{b}) Ratio of the computed surface area to that of a volume-equivalent sphere as a function of $\mathit{Ar}$. The analytical prediction (\ref{eq:4.6}) based on the spherical-cap model is shown for comparison (dotted line). (\textit{c}) Time evolution of $\mathit{Ar}(t)$ for varying $\mathit{Re}_0$, plotted against $(1-t/t_f)$. The scaling law $(1-t/t_f)^{-1/2}$ is included as a reference.}
  \label{fig:spherical-cap}
\end{figure}

To identify the origin of the discrepancy at moderate-to-high $\mathit{Re}_0$, \rec{we revisit the estimate $A(t) \sim V^{2/3}(t)$}, which implicitly assumes that the solid maintains a spherical morphology throughout melting \citep{yang2024shape}. However, the flow–melting interactions documented in \S~\ref{sec:melt-dynam-flow} demonstrate that for $\mathit{Re}_0 \gtrsim 150$ the rear surface flattens under the action of wake recirculation, yielding a geometry that more closely resembles a spherical cap: a rounded front with an approximately planar back, as sketched in figure~\ref{fig:spherical-cap}(\textit{a}). The instantaneous aspect ratio is therefore defined as $\mathit{Ar}(t) = l_b/l_a$, where $l_a$ and $l_b$ denote the streamwise and transverse extents in the picture, respectively. To capture the corresponding change in surface area, we compare the spherical-cap approximation with the volume-equivalent sphere. 
\rec{We define the area ratio $\mathcal{R} = A_{\mathrm{cap}}/A_{\mathrm{sph}}$, where $A_{\mathrm{cap}}$ represents the total surface area of the spherical-cap geometry, comprising both the rounded front interface and the flattened rear interface, as illustrated in figure~\ref{fig:spherical-cap}(\textit{a}), and $A_{\mathrm{sph}}$ denotes the surface area of the volume-equivalent sphere.}
A simple geometrical argument yields
\begin{align}\label{eq:4.6}
    \mathcal{R}(\mathit{Ar}) & =
    \begin{cases}
        \frac{2\mathit{Ar}-1}{(3\mathit{Ar}-2)^{2/3}}, & 2\geq \mathit{Ar} \geq 1 ,\\
        \frac{\frac{\mathit{Ar}^2}{2}+1}{(\frac{3}{4}\mathit{Ar}^2+1)^{2/3}}, & \mathit{Ar} > 2,
    \end{cases}
\end{align}
from the expression of the surface area of spherical caps with different aspect ratio \citep{polyaninHandbookMathematicsEngineers2006}. Figure~\ref{fig:spherical-cap}(\textit{b}) compares this expression with the numerically computed area ratios, confirming the validity of the spherical-cap approximation across $\mathit{Re}_0 \geq 150$. The results demonstrate that $A(t)$ cannot be approximated by $D_e^2(t)$ alone but requires explicit dependence on $\mathit{Ar}(t)$. The temporal evolution of $\mathit{Ar}(t)$ is shown in figure~\ref{fig:spherical-cap}(\textit{c}). Remarkably, for $\mathit{Re}_0 \in [150,1000]$ the aspect ratio collapses onto a universal scaling law,
\begin{equation}\label{eq:ar-law}
  \mathit{Ar}(t) \approx (1-\frac{t}{t_f})^{-1/2} \quad \text{for}~\mathit{Re}_0\in [150:1000],
\end{equation}
which directly links shape evolution to the melt progression. This scaling will be incorporated into an improved prediction for $V(t)$.

Given the self-similar nature of the system with respect to $\mathit{Re}_0$, we introduce the normalized variables $\hat{t} = t/t_f$ and $\hat{V} = V(t)/V_0$. Using the scaling $\bar{v}_{\mathit{\Gamma}} \sim (1-\hat{t})^{-1/3}$, the governing system for predicting the volume evolution can be written as
\begin{equation}\label{eq:further-pred-init}
  \hat{V}(\hat{t}=0) = 1,
\end{equation}
\begin{equation}\label{eq:further-pred-derivation}
  \frac{\mathrm{d}\hat{V}}{\mathrm{d}\hat{t}} = -\hat{\bar{v}}_{\mathit{\Gamma}}(\hat{t})\hat{A}_{\mathrm{cap}}(\hat{t}) = -K (1-\hat{t})^{-1/3}\mathcal{R}(\mathit{Ar}(\hat{t}))(6\sqrt{\pi}\hat{V})^{2/3},
\end{equation}
\begin{equation}\label{eq:further-pred-ar}
  \mathit{Ar}(\hat{t}) = (1-\hat{t})^{-1/2}
\end{equation}
\begin{equation}\label{eq:further-pred-final}
   \hat{V}(\hat{t}=0.9) = 0.01,
\end{equation}
where $\hat{\bar{v}}_{\mathit{\Gamma}}(\hat{t})$ denotes the rescaled global melting rate (normalized by the characteristic length $(6V_0/\pi)^{1/3}$ and time $t_f$), $K$ is a constant, and $(6\sqrt{\pi}\hat{V})^{2/3}$ corresponds to the surface area of a sphere of volume $\hat{V}$. 
\rec{
The terminal condition (\ref{eq:further-pred-final}) is imposed when the remaining solid fraction reaches 1\%, owing to the aforementioned technical limitations in accurately representing the interface as the solid approaches complete melting.
}

Because of the nonlinear dependence of $\hat{A}_{\mathrm{cap}}(\hat{t})$ on $\hat{t}$, the governing equation does not lead to a tractable closed-form expression. Although a formal solution can be written in terms of an integral, its expression is unwieldy and offers little practical insight. For clarity and efficiency, the system is therefore integrated numerically using a forward Euler scheme with 1000 time steps, which was verified to ensure convergence. The constant $K$ is iteratively adjusted to satisfy the terminal condition (\ref{eq:further-pred-final}), yielding $K \approx 0.368$. The resulting prediction, shown as the dash–dotted line in figure~\ref{fig:volume}, matches the simulation data closely, particularly at moderate-to-high $\mathit{Re}_0$, where the old scaling (\ref{eq:classical-scaling-law-prediction}) systematically overestimates the melting rate. This improved agreement highlights the importance of incorporating instantaneous shape effects, represented by $\mathcal{R}(\hat{t})$, when predicting the global volume evolution of melting bodies.

\section{Forces and Torques}\label{sec:drag-force-lift}

We now examine the hydrodynamic forces and torques acting on the melting body. These quantities are evaluated using the sharp-interface method proposed by \citet{xue_three-dimensional_2023}, which reconstructs the liquid–solid boundary from piecewise-linear facets within the volume-of-fluid framework and enables accurate surface integration. The total hydrodynamic force is given by
\begin{equation}\label{eq:total-force}
  \bm{F} = -\int_{\mathit{\Gamma}} \bm{\Sigma}\cdot\bm{n}~\mathrm{d}\mathcal{A},
\end{equation}
where $\bm{\Sigma} = -p\mathbf{I} + \rho^\ell\nu\nabla\bm{u}$ is the stress tensor, $\rho$ is the fluid density, $\bm{n}$ is the outward unit normal to the solid surface, and $\mathrm{d}\mathcal{A}$ denotes the differential surface element reconstructed from the interface. The force may be decomposed into its streamwise (drag) and transverse (lift) components, defined as $F_D = \bm{F}\cdot\bm{e}_x$ and $F_L = \bm{F}\cdot\bm{e}_\eta$, respectively, with $\eta$ denoting the symmetry-breaking direction. To enable comparison across different regimes, we define the corresponding dimensionless coefficients $C_D$ and $C_L$ using the force dimension $\rho^\ell U_\infty^2 A_x(t)/2$, where $A_x(t)$ is the instantaneous projected frontal area of the body onto the $\eta$--$\zeta$ plane. Both coefficients can be further partitioned into pressure and viscous contributions, $C_{D,p}$ ($C_{L,p}$) and $C_{D,\mu}$ ($C_{L,\mu}$), which will be reported separately where relevant. At Reynolds numbers exceeding the first bifurcation threshold ($\mathit{Re}_0 > 212$), the loss of axi-symmetry in the wake also induces a torque on the body. In dimensional form, the torque vector is  $-\int_{\mathit{\Gamma}} \bm{r} \times (\bm{\Sigma}\cdot\bm{n})~\mathrm{d}\mathcal{A}$, where $\bm{r}$ denotes the position vector relative to the body’s mass centroid. The associated dimensionless torque coefficients are defined as $T_i = -\bm{e}_i \cdot \int_{\mathit{\Gamma}} \bm{r} \times (\bm{\Sigma}\cdot\bm{n})~\mathrm{d}\mathcal{A}/\frac14\rho^\ell U_\infty^2 A^{3/2}_x(t)$, with $i=x,\eta\,\mathrm{or}~\zeta$. All results presented in the following sections are reported in terms of these dimensionless force and torque coefficients.

\subsection{Drag force}\label{sec:drag-force-coeff}

\begin{figure}
  \centering
  \includegraphics[width=\textwidth]{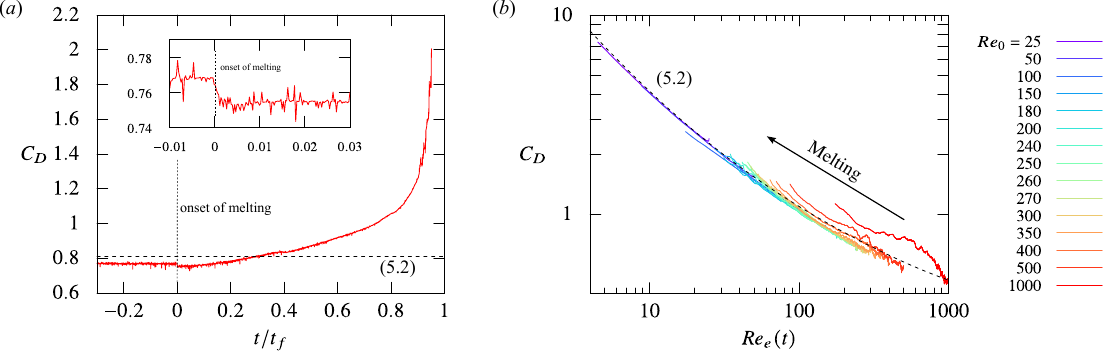}
  \caption{(\textit{a}) Time evolution of the drag coefficient $C_D$ for $\mathit{Re}_0=200$, with the inset providing a magnified view of the onset of melting. (\textit{b}) Variation of $C_D$ with the effective Reynolds number $\mathit{Re}_e(t)$ for different initial Reynolds numbers $\mathit{Re}_0$. In both panels, the dashed line denotes the empirical correlation for quasi-steady drag of rigid spheres given by (\ref{eq:drag-emprical}).}
  \label{fig:drag-force}
\end{figure}

Figure~\ref{fig:drag-force}(\textit{a}) presents the transient drag response for the case $\mathit{Re}_0 = 200$. The dashed line indicates the well-established empirical correlation for rigid spheres \citep{clift2005bubbles},
\begin{equation}\label{eq:drag-emprical}
     C_D^H = \frac{24}{\mathit{Re}} ( 1 + 0.15 \mathit{Re}^{0.687} ) + \frac{0.42}{1 + 4.25 \times 10^4 \mathit{Re}^{-1.16}},
\end{equation}
valid over $0 < \mathit{Re} < 3\times 10^5$. Before melting ($t \leq 0$), the drag force is in close agreement with (\ref{eq:drag-emprical}), as expected. Immediately after melting begins, however, $C_D(t)$ exhibits a short-lived but distinct drop for $0 < t/t_f < 0.002$, highlighted in the inset. This transient decrease will be analyzed in detail very soon. Following this brief interval, $C_D(t)$ increases monotonically as the effective Reynolds number $\mathit{Re}_e(t)$ decreases with the shrinking body size.

The overall dependence of $C_D(t)$ on $\mathit{Re}_e(t)$ is summarized in figure~\ref{fig:drag-force}(\textit{b}) for a range of $\mathit{Re}_0$. Because $\mathit{Re}_e(t)$ decreases with time, the trajectories should be interpreted from right to left. At moderate Reynolds numbers ($25 \leq \mathit{Re}_0 \leq 200$), the simulated drag coefficients collapse closely onto the rigid-sphere correlation (\ref{eq:drag-emprical}). In contrast, at higher $\mathit{Re}_0$ systematic deviations arise: the computed drag exceeds the empirical prediction. This discrepancy originates from the evolving morphology of the melting body. While at low-to-moderate $\mathit{Re}_0$ the body remains nearly spherical or slightly prolate, at higher $\mathit{Re}_0$ the interface develops a bluff, cup-cap-like geometry (\S~\ref{sec:axi-symm-melt}), which promotes earlier separation and enhanced pressure drag, thereby exceeding the rigid-sphere correlation.

Moreover, caution is required when interpreting the evolution of $C_D(t)$ as a function of $\mathit{Re}_e(t)$, since the body undergoes simultaneous variations in both size and shape. Such temporal changes can, in principle, induce unsteady hydrodynamic contributions through added-mass and history effects. To the best of our knowledge, no rigorous theory exists for the unsteady drag of a single melting or dissoluble body, even in the idealized case of a perfect sphere. Nevertheless, insight may be gained by analogy with isotropic vapor bubbles in superheated or subcooled liquids, as studied by \citet{legendre1998thermal}. Importantly, the added-mass coefficient depends only on body shape and not on the interfacial condition \citep{mougin2002generalized}. Thus, for a spherical body, whether rigid or slippery, the well known added-mass coefficient is $C_M = 1/2$. This equivalence strongly suggests that both bubble collapse and sphere melting experience the same added-mass correction. For a sphere with time-varying diameter, the added-mass correction to the drag coefficient may be written as $C_I(t) = 8C_M\mathcal{U}(t)$, where $\mathcal{U}(t) = D_e'(t)$ is the ratio of the interfacial radial velocity to the free-stream velocity. \reb{In the present context, $\mathcal{U}(t)$ can be estimated from the Stefan condition (\ref{eq:interface-condition-2}) as $\mathcal{U}(t) \approx -\mathit{St}(\mathit{Re}_0\mathit{Pr})^{-1}\delta_\theta^{-1}(t)$. With $\delta_\theta \approx 1.13\mathit{Pr}^{-1/2}\mathit{Re}_0^{-1/2}D_e^{1/2}(t)$ \citep{schlichting_boundary-layer_2000}, this scaling reduces to $\mathcal{U}(t) \approx -0.885St \mathit{Pr}^{-1/2}\mathit{Re}_e^{-1/2}(t)$. Substituting $\mathit{St} = 0.25$ and $\mathit{Pr}=7$, we obtain $\mathcal{U}(t) \approx -0.0836\mathit{Re}_e^{-1/2}(t)$, and therefore $C_I(t) \approx -0.334\mathit{Re}_e^{-1/2}(t)$. This correction is modest: $C_I \approx -0.106$ for $\mathit{Re}_e=10$, $-0.0236$ for $\mathit{Re}_e=200$, and $-0.0106$ for $\mathit{Re}_e=1000$. }    

\begin{figure}
    \centering
    \includegraphics[width=0.9\textwidth]{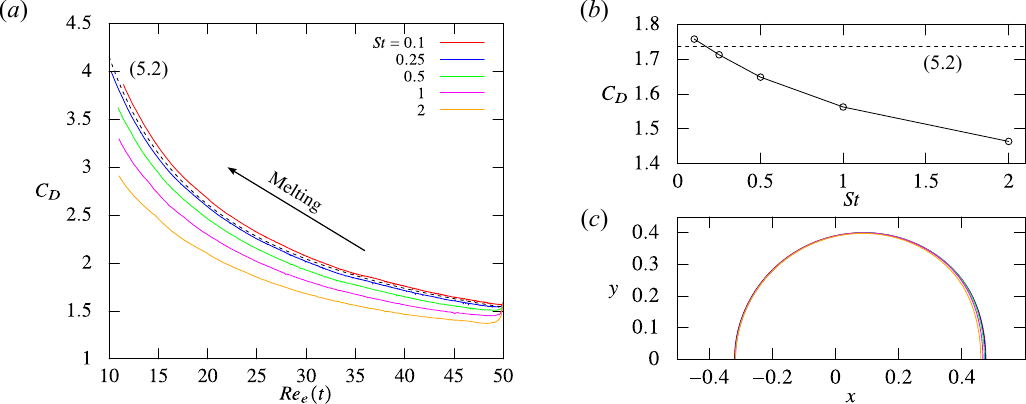}
    \caption{Effect of Stefan number $\mathit{St}$ on the temporal evolution of drag at $\mathit{Re}_0=50$. (\textit{a}) Drag coefficient $C_D$ as a function of effective Reynolds number $\mathit{Re}_e(t)$ for varying $\mathit{St}$. (\textit{b}) Dependence of $C_D$ on $\mathit{St}$ at fixed $\mathit{Re}_e=40$. (\textit{c}) Mid-plane interface ($z=0$) at $\mathit{Re}_e=40$ for different $\mathit{St}$, showing negligible morphological differences. Panels (\textit{a}) and (\textit{c}) share the same legend.}
    \label{fig:added-mass-st}
\end{figure}

In all cases, the added-mass correction remains below $4\%$ of the quasi-steady drag coefficient (\ref{eq:drag-emprical}), and can therefore be neglected for the present simulations at $\mathit{St}=0.25$. At higher Stefan numbers, however, the increased interfacial recession velocity suggests that added-mass effects may become significant. This expectation is confirmed in figure~\ref{fig:added-mass-st}. Panel (\textit{a}) shows the evolution of $C_D(t)$ versus effective Reynolds number $\mathit{Re}_e(t)$ at fixed $\mathit{Re}_0=50$ for $\mathit{St}$ between $0.1$ and $2$. A clear trend emerges: larger $\mathit{St}$ systematically reduces $C_D(t)$ at identical $\mathit{Re}_e(t)$. Panel (\textit{b}) quantifies this effect, showing that at $\mathit{Re}_e=40$ the instantaneous drag decreases from $1.71$ at $\mathit{St}=0.25$ to $1.47$ at $\mathit{St}=2$, consistent with the scaling $\mathcal{U}(t)\sim -St$ derived above. Since figure~\ref{fig:added-mass-st}(\textit{c}) demonstrates that interface morphology is nearly unchanged across cases, the drag reduction cannot be attributed to shape effects but instead reflects the growing role of added-mass contributions at higher melting rates.

As for the history force, \citet{legendre1998thermal} and \citet{magnaudet1998viscous} derived analytical expressions for the history coefficient $C_H(t)$ for bubbles with time-varying radii in the creeping-flow limit. In this regime, $C_H(t)$ originates from the diffusion of vorticity around the bubble and becomes relevant only when $\mathit{Re}_e(t)\ll 1$ and $\mathcal{U}\mathit{Re}_e(t)\ll 1$. At higher Reynolds numbers ($\mathit{Re}_e(t)\gg 1$), the history effect reduces to a viscous correction consistent with viscous potential flow, scaling as $C_H(t) = 48/\mathit{Re}_e(t) - C_D(t)$. This correction is significant at very early times but decays rapidly as $\mathit{Re}_e$ grows. In the present study, $\mathit{Re}_e(t)\gg 1$ throughout most of the melting process, and although the viscous-potential assumption is not strictly valid for rigid solids, \citet{legendre1998thermal} showed that $C_H(t)$ is always smaller than the added-mass correction $C_I(t)$ for vaporizing bubbles. Hence, it is reasonable to conclude that the history force is negligible here as well. Taken together, both $C_I(t)$ and $C_H(t)$ make only minor contributions to the total drag, which explains why for $25 \leq \mathit{Re}_0 \leq 200$ the computed drag coefficient $C_D(t)$ follows the quasi-steady prediction (\ref{eq:drag-emprical}) so closely.

\begin{figure}
  \centering
  \includegraphics[width=\textwidth]{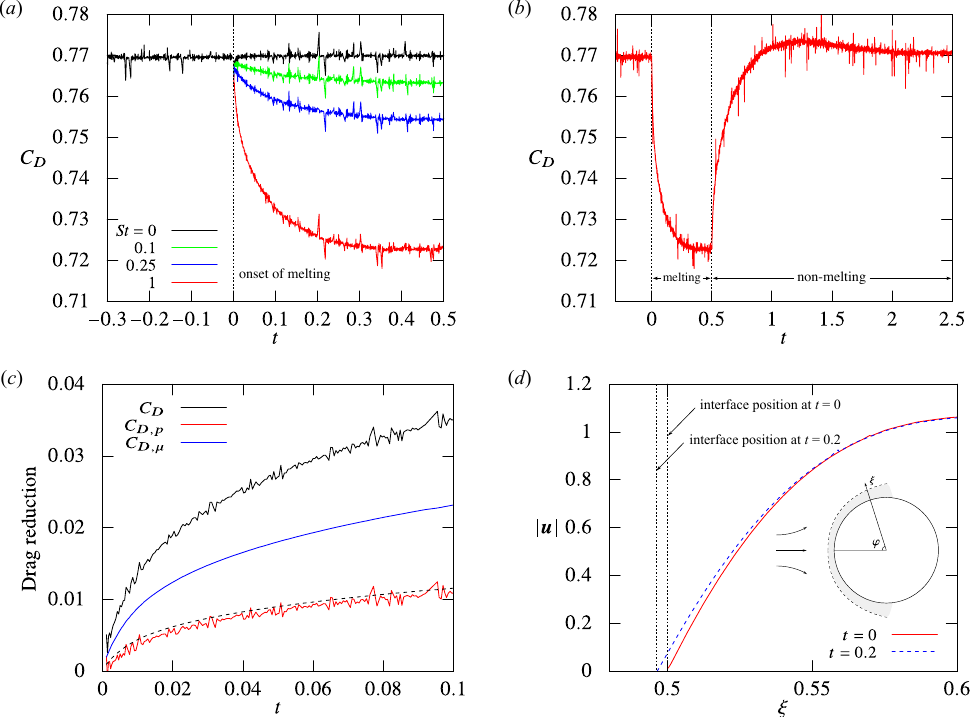}
   \caption{
   \reb{Quantitative results of the sudden drop in drag coefficient $C_D (t)$ immediately after melting begins, corresponding to $\mathit{Re}_0=200$.
(\textit{a}) Time evolution of $C_D (t)$ at different Stefan numbers, showing larger drag reductions at higher $\mathit{St}$.
(\textit{b}) Auxiliary test of the drag coefficient $C_D(t)$ at $\mathit{St}=1$, in which melting is switched on at $t=0$ and subsequently switched off at $t=0.5$. Once melting ceases, $C_D(t)$ returns to its pre-melting value, confirming that the observed drag drop is entirely attributable to the melting process.
(\textit{c}) Decomposition of drag reduction ($\Delta C_{D}$, black solid line) into pressure ($\Delta C_{D,p}$, red line) and viscous ($\Delta C_{D,\mu}$, blue line) contributions for $\mathit{St} = 1$, with dashed line showcasing $\Delta C_{D,\mu}/2$ for reference.
(\textit{d}) Velocity magnitude profiles $|\bm{u}|$ at $\varphi = 63^\circ$, comparing $t=0$ (red solid line) and $t=0.2$ (blue dashed line) for $\mathit{St}=1$. Surface recession suddenly thickens the boundary layer and thus reduces viscous shear.
}}
  \label{fig:force-onset-drop}
\end{figure}

We now return to the abnormal dip observed in figure~\ref{fig:drag-force}(\textit{a}) at the onset of melting, $t/t_f \to 0^+$, a closer examination reveals that $C_D(t)$ undergoes a sudden reduction immediately after melting begins. Figure~\ref{fig:force-onset-drop}(\textit{a}) shows this onset behaviour for $\mathit{Re}_0=200$ at different Stefan numbers, $\mathit{St}=0.1$, $0.25$, and $1$. The magnitude of the drop clearly intensifies with increasing $\mathit{St}$, from $\Delta C_D \approx 0.008$ at $\mathit{St}=0.1$ to $\Delta C_D \approx 0.048$ at $\mathit{St}=1$. This trend indicates that faster melting rates induce stronger drag reduction.
\reb{
 Additionally, to confirm that the sudden drag drop is solely caused by the melting process rather than any numerical artefact associated with activating the phase change, we conduct an auxiliary test at $\mathit{St}=1$. In this test, melting is switched on at $t=0$ and subsequently switched off at $t=0.5$. As shown in figure~\ref{fig:force-onset-drop}(\textit{b}), once melting is turned off, the reduced $C_D(t)$ rises back to its pre-melting value, as expected.
}
Classical diffusion-controlled history forces are unlikely to explain this effect in the $\mathit{Re}_e \gg 1$ regime \citep{magnaudet1998viscous}. Instead, as \citet{magnaudet1998viscous} reported for bubbles, when $\mathit{Re}_e\gg 1$ and $\mathcal{U}\mathit{Re}_e\gg 1$, surface recession alters the local radial velocity field, thereby modifying the viscous contribution to the drag. The effect strengthens with larger $\mathcal{U}$ (or $\mathit{St}$) and is most pronounced at the very onset of melting. This interpretation is confirmed in figure~\ref{fig:force-onset-drop}(\textit{c}), which decomposes the drag reduction (black line) for $\mathit{St}=1$ into its viscous (blue line) and pressure (red line) components. The two contributions satisfy $\Delta C_{D,\mu} \approx 2\Delta C_{D,p}$, consistent with the well-known balance that viscous pressure is approximately half of the viscous drag \citep{legendre1998lift}. This demonstrates that the observed drag reduction originates from viscous, rather than inertial effects. Further evidence is provided in figure~\ref{fig:force-onset-drop}(\textit{d}), where the velocity profiles inside the boundary layer at $\varphi=63^\circ$ show that surface recession thickens the viscous boundary layer between $t=0$ and $t=0.2$, thereby reducing $\partial |\bm{u}|/\partial \xi$ at the wall and hence the viscous shear stress. This mechanism is consistent with the experimental observations of \citet{vakarelski2015drag}, who reported drag reduction during the initial stages of sphere melting and attributed it to effective boundary-layer thickening due to interfacial recession.

\subsection{Lift force}\label{sec:lift-force}

\begin{figure}
  \centering
  \includegraphics[width=0.99\textwidth]{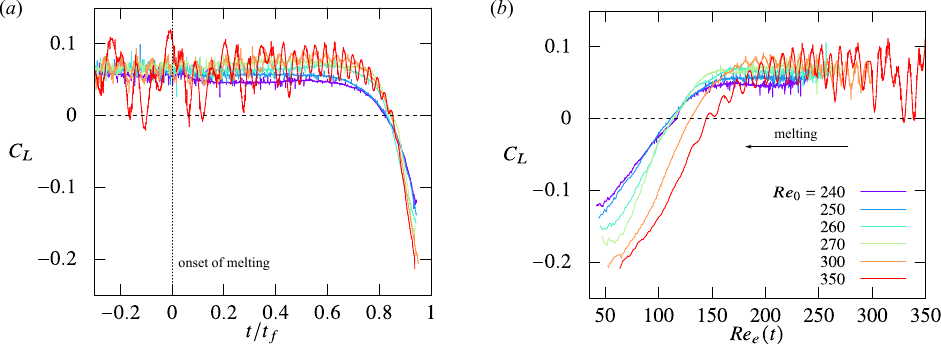}
  \caption{
  Evolution of the lift coefficient $C_L$ in the \textit{steady-planar-symmetric} and \textit{periodic-planar-symmetric melting regimes} ($212 < \mathit{Re}_0 < 355$).
(\textit{a}) $C_L$ as a function of dimensionless time $t/t_f$, showing a sharp decline near $t/t_f \approx 0.7$ and eventual reversal at late times.
(\textit{b}) $C_L$ versus effective Reynolds number $\mathit{Re}_e$, illustrating that larger $\mathit{Re}_0$ accelerates the decline and reversal of lift due to stronger asymmetry in rear-interface melting. 
}
\label{fig:lift-force}
\end{figure}

We now turn to the lift force acting on the melting sphere, focusing on the \textit{steady-planar-symmetric} and \textit{periodic-planar-symmetric melting regimes} ($212 < \mathit{Re}_0 < 355$). Figure~\ref{fig:lift-force} summarizes the evolution of the lift coefficient $C_L$, plotted against both dimensionless time $t/t_f$ and effective Reynolds number $\mathit{Re}_e(t)$. In the absence of melting ($t/t_f < 0$), the lift response agrees with prior characterizations of the wake: $C_L$ is steady for $212 < \mathit{Re}_0 \lesssim 270$, but becomes oscillatory for $270 \lesssim \mathit{Re}_0 \lesssim 300$, consistent with the transition boundary reported by \citet{johnson_flow_1999}.

Once melting commences, the early-to-intermediate stages ($t/t_f <  0.7$) retain this pre-melting behavior, indicating that modest interfacial recession does not immediately reorganize the lift dynamics. Then after that ($t/t_f >  0.7$), $C_L(t)$ undergoes a sharp decline when the morphology change marks the transition from a spherical-cap-like to a cup-cap geometry, as previously displayed in figure~\ref{fig:spherical-cap}(\textit{a}). Physically, it corresponds to accelerated melting of the lower rear interface and the subsequent growth of the lower spiral, which restores approximate vorticity balance across the wake and suppresses the net lift force. For later times ($t/t_f > 0.8$), $C_L$ not only vanishes but eventually reverses sign for all cases considered, implying that a freely translating body would migrate from the positive to the negative $\eta$-direction. The $C_L$–$\mathit{Re}_e$ representation in figure~\ref{fig:lift-force}(\textit{b}) highlights that the decay and reversal occur earlier for larger $\mathit{Re}_0$. This enhanced sensitivity arises because higher $\mathit{Re}_0$ strengthens the asymmetry between upper and lower melting rates at the rear interface (cf. figure~\ref{fig:plane-back-planar-interface}), accelerating the reorganization of the wake.

\begin{figure}
  \centering
  \includegraphics[width=\textwidth]{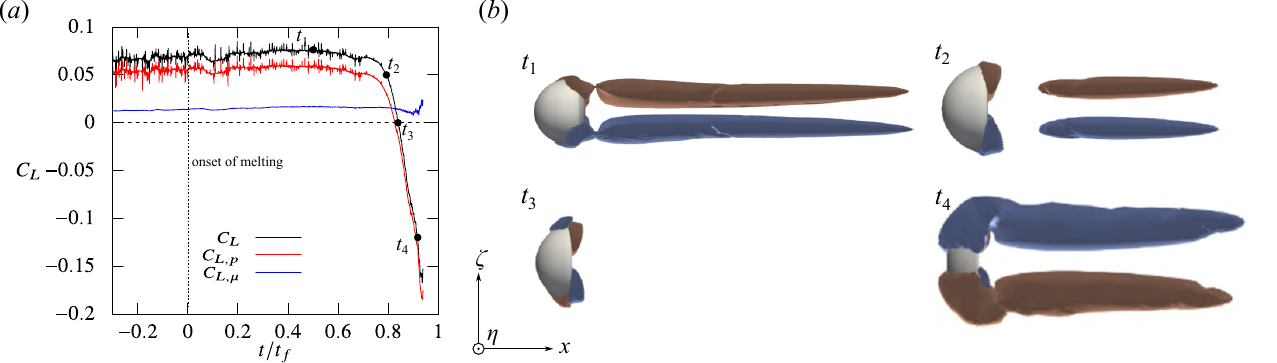}
  \caption{Lift-force decomposition and streamwise vorticities at $\mathit{Re}_0=270$.
(\textit{a}) Time evolution of the lift coefficient $C_L$ its pressure ($C_{L,p}$) and viscous ($C_{L,\mu}$) contributions, showing that the reversal of $C_L$ is controlled almost entirely by the pressure component.
(\textit{b}) Iso-contours of streamwise vorticity $\omega_x$ on the $x$--$\zeta$ plane (red: $\omega_x=0.25$, blue: $\omega_x=-0.25$) at four instants $t_1$–$t_4$ marked in panel (\textit{a}). The exchange in dominance between upper and lower spirals reverses the sign of $\omega_x$ and hence the direction of lift.}
\label{fig:re-270-lift-force}
\end{figure}

To further clarify the mechanism underlying lift reversal in the late stage of melting, we analyze in detail the case $\mathit{Re}_0=270$, as presented in figure~\ref{fig:re-270-lift-force}(\textit{a}). The lift coefficient $C_L(t)$ is decomposed into its pressure and viscous components, $C_{L,p}(t)$ and $C_{L,\mu}(t)$. \reb{The viscous contribution remains nearly constant throughout the process, indicating that shear stresses play only a secondary role in the lift reversal.} By contrast, the pressure component dominates and is directly responsible for the reversal of $C_L$. The inertial origin of this effect can be traced to the relative intensities of the lower and upper spirals, which reverse the sign of $\omega_\zeta^{\eta > 0} - \omega_\zeta^{\eta < 0}$, and subsequenty, such reversal also change the orientation of the produced counter-rotating streamwise vortices in the wake, thus flips the sign of $C_{L,p}$. This process is visualized in figure~\ref{fig:re-270-lift-force}(\textit{b}), which shows iso-contours of streamwise vorticity $\omega_x=\pm 0.25$ at representative time instants ($t_1$–$t_4$ shown in panel \textit{a}). The reversal of lift coincides precisely with a change in the dominant sign of $\omega_x$, confirming that the transverse force arises directly from the counter-rotating vortex pair in the wake, as anticipated in the classical picture of bluff-body lift generation \citep{lighthill1956drift}. Since $\omega_x$ itself is produced by stretching and tilting of azimuthal vorticity $\omega_\zeta$, the exchange of dominance between the upper and lower spirals naturally reorganizes the wake structure and drives the observed lift reversal.

\subsection{Torque}\label{sec:torque}

We now examine the torques acting on the melting sphere, since a freely descending body would rotate in response, thereby introducing an additional degree of freedom that could modify the melting dynamics. For clarity, we focus on the representative case $\mathit{Re}_0=270$, noting that other Reynolds numbers exhibit qualitatively similar behaviour.

\begin{figure}
    \centering
    \includegraphics[width=0.9\textwidth]{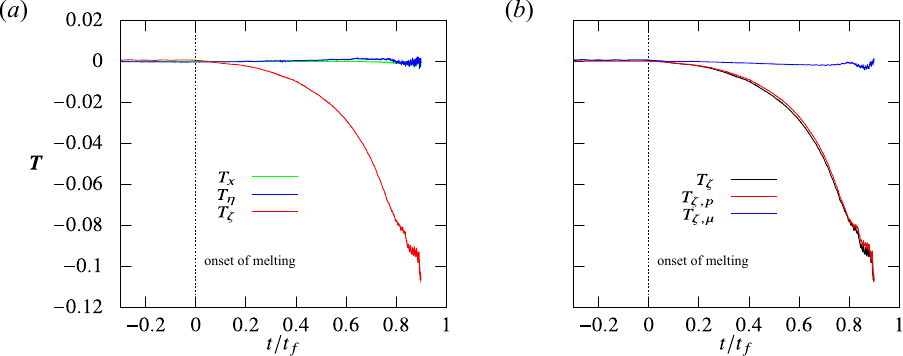}
    \caption{Torque evolution experienced by the melting sphere at $\mathit{Re}_0=270$.
(\textit{a}) Time evolution of torque coefficients $T_x$, $T_\eta$ and $T_\zeta$. Only the $\zeta$-component becomes significant over time, while $T_x$ and $T_\eta$ remain negligible.
(\textit{b}) Decomposition of $T_\zeta$ into pressure ($T_{\zeta,p}$) and viscous ($T_{\zeta,\mu}$) contributions, showing that the torque is almost entirely pressure-driven.}
    \label{fig:re-270-torque-1}
\end{figure}

Figure~\ref{fig:re-270-torque-1}(\textit{a}) shows that only the torque component $T_\zeta$ becomes dynamically relevant, growing noticeably as soon as the melt starts $t/t_f > 0$, while $T_x$ and $T_\eta$ remain negligible. Decomposition $T_\zeta$ into pressure and viscous contributions, figure~\ref{fig:re-270-torque-1}(\textit{b}) reveals that $\zeta$-component torque is almost entirely pressure-driven. Thus, the asymmetric pressure distribution generates a steadily increasing anticlockwise torque about the $\zeta$-axis, perpendicular to the $C_D$--$C_L$ plane. Examination of instantaneous shapes (see figure~\ref{fig:plane-3D-melting-processes}) shows that this anticlockwise torque tends to reorient the inclined rear face of the body towards a vertical configuration. If free to rotate, the body would oscillate about this orientation, effectively homogenising the melting rate over its surface. This behaviour is consistent with the experiments of \citet{machicoane2013melting}, who reported that freely moving ice spheres in turbulent flows preserved a nearly spherical shape for long durations owing to their ability to rotate.

\begin{figure}
    \centering
    \includegraphics[width=\textwidth]{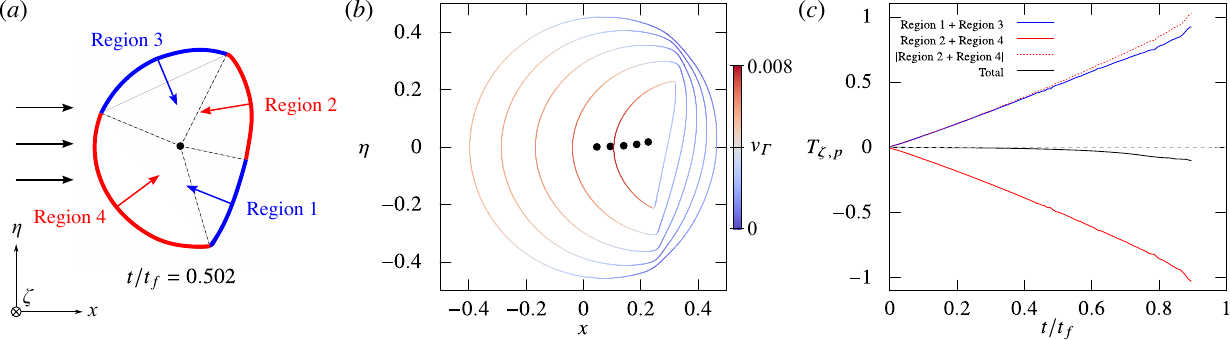}
    \caption{Clarification of the origin of the anticlockwise torque acting on the body for $\mathit{Re}_0=270$. (\textit{a}) Illustration of the four regions of the interface at the symmetry plane ($x$--$\eta$) at $t/t_f=0.502$, where region 1 and region 3 generate clockwise moments (blue portions) while region 2 and region 4 (red portions) generate anticlockwise moments. The black filled circle represents the mass centre. It's observed that region~4 directly faces the oncoming flow, resulting in elevated pressure relative to region~3. (\textit{b}) Time evolution of the interface at the symmetry plane ($x$--$\eta$)  and the mass centre represented by the black filled circle, which confirms that melting drive a northeastward shift of the mass centre. (\textit{c}) Time evolution of the pressure component of the torque $T_{\zeta,p}$, with the net contribution from region 1 + region 3 (blue) and region 2 + region 4 (red). For ease of comparison, the red dotted curve represents the absolute values of the red solid curve.}
    \label{fig:re-270-torque-2}
\end{figure}

The origin of this anticlockwise torque is clarified in figure~\ref{fig:re-270-torque-2}(\textit{a}), where the body surface is partitioned into four regions at $t/t_f=0.502$ in an $x$--$\eta$ projection. Regions~2 and 4 generate anticlockwise moments (red portins), while regions~1 and 3 produce clockwise moments (blue portions). Owing to the inclination of the body, regions~2 and 4 possess larger effective surface areas. Moreover, region~4 directly faces the oncoming flow, resulting in elevated pressure relative to region~3. These geometric and hydrodynamic factors combine to favour anticlockwise torque. The time evolution of centroid position in figure~\ref{fig:re-270-torque-2}(\textit{b}) confirms that melting drives a northeastward shift, further amplifying contributions from regions~2 and 4. Quantitative evidence is given in figure~\ref{fig:re-270-torque-2}(\textit{c}), which compares net contributions: anticlockwise torques dominate progressively, yielding the observed monotonic growth of $T_{\zeta,p}$ and hence $T_\zeta$.

\section{Effect of buoyancy force}\label{sec:buoyancy}

We next examine the role of thermal buoyancy in modifying the melting dynamics of spheres. In the governing equations, buoyancy enters through the body-force term $-\mathit{Ri}\,\theta\,\bm{e}_i$ in (\ref{eq:ns1}) \rec{under the Boussinesq approximation}, introducing mixed-convection effects due to the combined action of forced and natural convection. 
\rec{
Here, particular care must be taken regarding the correlation between fluid density and temperature. A linear relation is commonly adopted to represent thermal expansion. However, for an ice–water system, a more refined quadratic density–temperature relation more faithfully captures the density anomaly of water, whose density reaches a maximum near $4~^\circ\mathrm{C}$, particularly when the inflow temperature is below $6~^\circ\mathrm{C}$ \citep{weadyAnomalousConvectiveFlows2022}.
The aim of this section is not to carry out an exhaustive parametric study of mixed-convection melting, but rather to illustrate the qualitative influence of buoyancy when it becomes comparable to inertia.
Accordingly, we adopt the linear density–temperature relation, which is broadly applicable to generic liquids, including water at $20^\circ\mathrm{C}$, the reference temperature at which the dimensionless parameters of this study are defined. We further show that, although the density anomaly introduces some local quantitative differences, the overall melting behaviour and flow regimes at this reference temperature remain largely consistent between the linear and quadratic density–temperature relations.
Moreover, to explicitly address the density anomaly of water for varying ambient temperature, a representative case using the quadratic relation is additionally examined and discussed at the end of this section.
}
We focus on two representative configurations: (i) Gravity aligned with the streamwise direction ($\bm{e}_i=\bm{e}_x$), were $\mathit{Ri}\in[-0.5,0.5]$ distinguishes ascending ($\mathit{Ri}>0$) from descending ($\mathit{Ri}<0$) motions of a melting sphere in water; (ii) gravity perpendicular to the streamwise direction ($\bm{e}_i=\bm{e}_y$), while in this orientation, buoyancy drives a cross-stream motion, and we consider $\mathit{Ri}=-0.1$ and $-0.5$, corresponding to horizontally translating spheres in water subject to increasing buoyancy influence. These two cases are analysed separately in \S\S~\ref{sec:gravity-aligned} and \ref{sec:gravity-perpendicular}, respectively.
For reference, a Richardson number of $\mathit{Ri} = \pm 0.5$ corresponds to a sphere of diameter $D_0 = 3~\mathrm{cm}$ translating at $U_\infty = 0.032~\mathrm{m/s}$ in water with an ambient temperature difference of $\Delta T = 20^\circ \mathrm{C}$.  

\subsection{Vertically translating sphere parallel to gravity}
\label{sec:gravity-aligned}

We first consider a melting sphere translating vertically in warm water. In this configuration, the colder melt released from the solid is subject to buoyancy acceleration in the streamwise direction. Depending on the orientation of the translation relative to gravity, two distinct flow responses arise. When the sphere rises against gravity ($\mathit{Ri} > 0$), The buoyant melt fluid is entrained downstream along the free-stream direction. This assisting buoyancy enhances momentum transport in the boundary layer, stabilises its evolution, and delays separation. By contrast, when the sphere descends with gravity ($\mathit{Ri} < 0$), the cold melt sinks counter to the free stream. This opposing buoyancy destabilises the boundary layer, promotes earlier separation, and strengthens the wake recirculation \citep{kotoucTransitionTurbulenceWake2009}. These competing tendencies substantially alter the interfacial evolution compared with the purely forced-convection case.

\begin{figure}
    \centering
    \includegraphics[width=\textwidth]{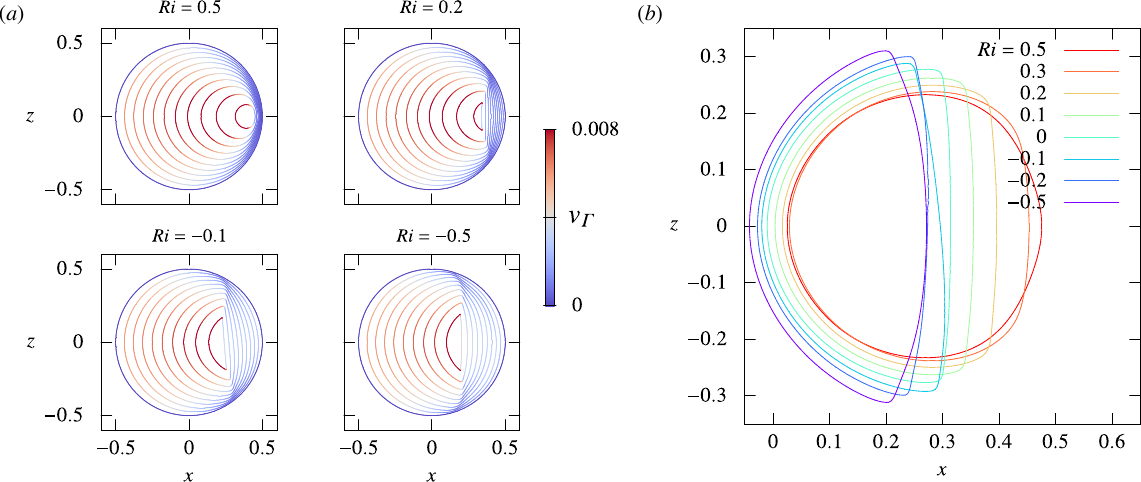}
    \caption{Influence of the buoyancy on melting dynamics when the gravity is aligned with the streamwise direction, the initial Reynolds number is maintained at $\mathit{Re}_0=200$. (\textit{a}) Time evolution of the interfacial shape, coloured by the local melting rate $v_\mathit{\Gamma}$, on the mid-plane at $y=0$ for four representative Richardson numbers, respectively of $\mathit{Ri}=0.5, 0.2, -0.1$ and $-0.5$. The temporal sampling interval is fixed at $\Delta t =10$ for all cases. (\textit{b}) Superposition of the mid-plane interface ($y=0$) when the remaining solid volume reaches $V(t)/V_0=0.1$, highlighting the contrasting influence of stabilising ($\mathit{Ri}>0$) and destabilising ($\mathit{Ri}<0$) buoyancy.}
    \label{fig:ri-slice}
\end{figure}

To isolate buoyancy effects, we fix the initial Reynolds number at $\mathit{Re}_0=200$ and vary the Richardson number in the range $\mathit{Ri}\in[-0.5,0.5]$. Figure~\ref{fig:ri-slice}(\textit{a}) presents instantaneous cross-sections of the melting body on the vertical mid-plane at $y=0$, coloured by the local melting rate $v_\mathit{\Gamma}$, for four representative cases: $\mathit{Ri}=-0.5, -0.1, 0.2$ and $0.5$. The influence of buoyancy on the separation location is readily apparent. For example, at $\mathit{Ri}=0.5$ the assisting buoyancy delays separation considerably, and the body retains an ellipsoidal morphology. In contrast, at $\mathit{Ri}=-0.5$ the opposing buoyancy enhances the rear recirculation, producing an exaggeratedly flattened back interface. A more quantitative comparison is given in figure~\ref{fig:ri-slice}(\textit{b}), which overlays the two-dimensional interfaces corresponding to different $\mathit{Ri}$ when the remaining solid volume has decayed to $V(t)/V_0=0.1$. The contrasting buoyancy effects are evident: negative $\mathit{Ri}$ yields strongly flattened rears, while positive $\mathit{Ri}$ produces more rounded geometries.

\begin{figure}
    \centering
    \includegraphics[width=\textwidth]{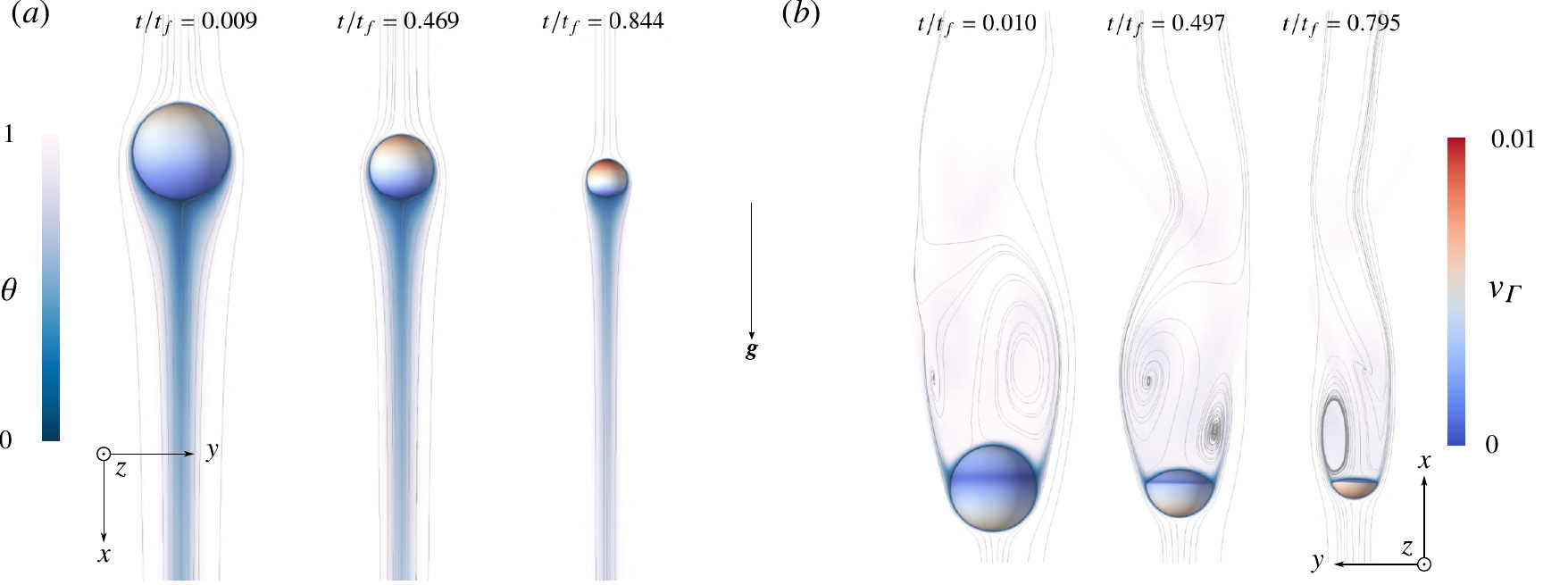}
    \caption{Melting dynamics of a sphere at $\mathit{Re}_0=200$ with gravity parallel to the streamwise direction. (\textit{a}) Assisting buoyancy case, $\mathit{Ri}=0.5$. (\textit{b}) Opposing buoyancy case, $\mathit{Ri}=-0.5$. Shown are snapshots of the temperature field $\theta$ and streamlines on the mid-plane at $z=0$, together with the 3D interface coloured by the local melting rate $v_\mathit{\Gamma}$, at three representative instants: $t=1, 50, 80$ ($t/t_f=0.009,0.469,0.844$ for $\mathit{Ri}=0.5$; $t/t_f=0.010,0.497,0.795$ for $\mathit{Ri}=-0.5$)}
    \label{fig:ri-melting-processes}
\end{figure}

Furthermore, even in the absence of phase change, i.e. for an adiabatically heated sphere, \citet{kotoucTransitionTurbulenceWake2009} demonstrated that buoyancy strongly alters wake dynamics, giving rise to distinct vortical regimes as $\mathit{Ri}$ varies. More recently, \citet{Li_Jiang_Xu_Zhao_2025} extended this analysis to freely translating spheres, identifying attached flow at $\mathit{Ri}=0.5$, axi-symmetric wakes at $\mathit{Ri}=0.1$ and $0.2$, steady-planar-symmetric flow at $\mathit{Ri}=-0.1$, quasi-periodic vortex shedding at $\mathit{Ri}=-0.2$, and fully chaotic flow at $\mathit{Ri}=-0.5$. These canonical regimes are faithfully reproduced in the present simulations of melting spheres. Figure~\ref{fig:ri-melting-processes} illustrates the contrasting cases of $\mathit{Ri}=0.5$ and $\mathit{Ri}=-0.5$: the assisting buoyancy stabilises the wake into an attached-flow configuration, whereas the opposing buoyancy promotes strong recirculation and chaotic vortex shedding. The same sequence of transitions is also evident in figure~\ref{fig:ri-slice}, which compares mid-plane interfacial morphologies across different $\mathit{Ri}$. These results confirm that the stabilising and destabilising effects of buoyancy on wake topology remain the dominant organising principle, even when coupled to phase change.

\subsection{Horizontally translating sphere perpendicular to gravity}\label{sec:gravity-perpendicular}

\begin{figure}
    \centering
    \includegraphics[width=\textwidth]{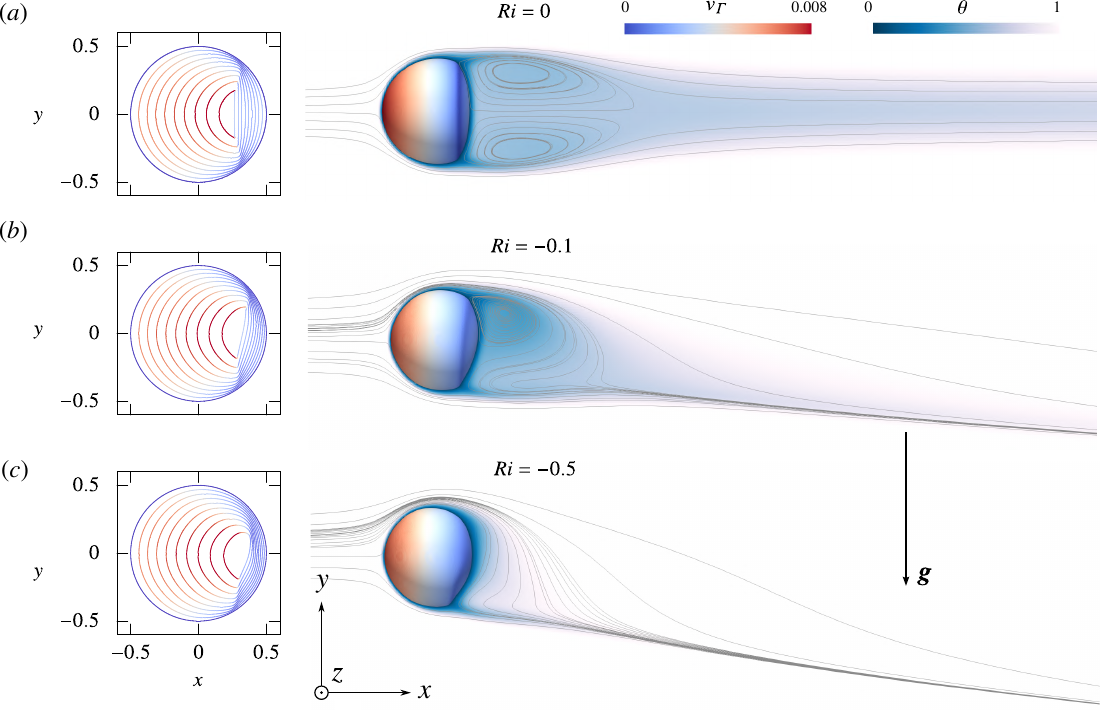} 
    \caption{Melting dynamics of a horizontally translating sphere at $\mathit{Re}_0=200$ with gravity acting perpendicular to the streamwise direction. Cases shown: (\textit{a}) $\mathit{Ri}=0$, (\textit{b}) $\mathit{Ri}=-0.1$, (\textit{c}) $\mathit{Ri}=-0.5$. Left panels: interfacial evolution on the mid-plane at $z=0$, coloured by the local melting rate $v_\mathit{\Gamma}$, with fixed temporal spacing $\Delta t=10$. Right panels: snapshots of the temperature field $\theta$ and streamlines on the mid-plane at $z=0$, together with the 3D interface coloured by $v_\mathit{\Gamma}$ at $t=40$ ($t/t_f = 0.397,0.400,0.408$ for $\mathit{Ri}=0,-0.1,-0.5$). Increasingly negative $\mathit{Ri}$ produces stronger wake asymmetry through stabilisation of the upper and destabilisation of the lower boundary layer.}
    \label{fig:ri-vert-melting-processes}
\end{figure}

We now consider the case where gravity acts perpendicular to the sphere’s translational direction, corresponding to horizontal motion in warm water. In this configuration, the melt-induced cold fluid is always deflected downward ($-y$ direction) by buoyancy, ensuring that the Richardson number remains negative ($\mathit{Ri}<0$). Unlike the vertical case, buoyancy here does not simply assist or oppose the streamwise flow; instead, it perturbs the wake to select a preferred reflectional plane of bifurcation. As a result, flow asymmetry necessarily develops in the $x$–$y$ plane. To isolate this effect, we fix the initial Reynolds number at $\mathit{Re}_0=200$ and examine two representative values, $\mathit{Ri}=-0.1$ and $-0.5$, with $\mathit{Ri}=0$ included as a reference.

The results, shown in figure~\ref{fig:ri-vert-melting-processes}, demonstrate that buoyancy acts asymmetrically on the boundary layer: the downward motion of the cold melt stabilises the upper hemisphere ($y>0$), delaying separation, while destabilising the lower hemisphere ($y<0$) and promoting earlier separation. Consequently, the wake and melting dynamics become increasingly asymmetric as $|\mathit{Ri}|$ increases. At $\mathit{Ri}=0$, the wake remains axisymmetric throughout the melting process. At $\mathit{Ri}=-0.1$, however, a tilt develops in the wake, producing unequal spirals and a rear interface inclined backward. This configuration closely resembles the \textit{steady-planar-symmetric regime} observed at $\mathit{Re}_0=270$ (see figure~\ref{fig:plane-3D-melting-processes}(\textit{c})). Under stronger buoyancy forcing ($\mathit{Ri}=-0.5$), the downward convection nearly suppresses the lower spiral, resulting in a highly asymmetric wake. At later times (left panel of figure~\ref{fig:ri-vert-melting-processes}(\textit{c})), the interface develops a rounded upper edge and a sharp lower edge, consistent with the experimental observations of \citet{hao2001melting} due to the asymmetric flow separations.

\begin{figure}
        \centering
        \includegraphics[width=0.9\textwidth]{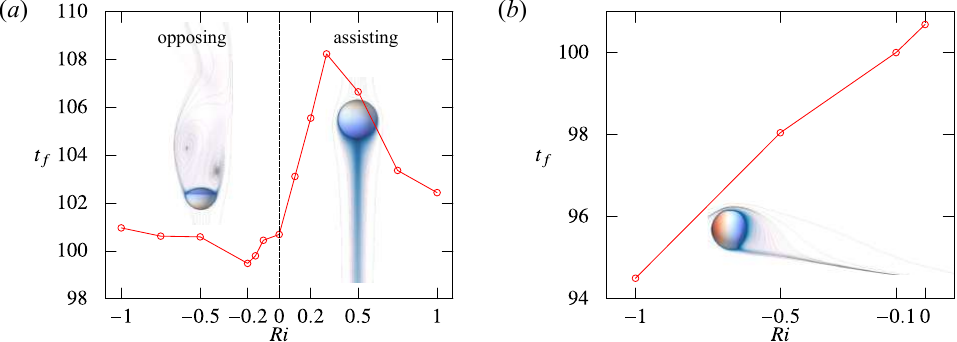}
        \caption{
    \rea{Dependence of the complete melting time $t_f$ on the Richardson number $\mathit{Ri}$ at fixed $\mathit{Re}_0=200$.
    (\textit{a}) Vertical translation (gravity parallel to the streamwise direction): within $-0.2 \leq \mathit{Ri} \leq 0.3$, $t_f$ increases with $\mathit{Ri}$, corresponding to a transition from a cup–cap shape to a prolate shape. For $\mathit{Ri} > 0.3$, the wake is increasingly suppressed, and melting is enhanced by progressively tighter attached flow. For $\mathit{Ri} \leq -0.5$, buoyancy induces chaotic recirculation, and the global melting rate evolution reaches a plateau.
    (\textit{b}) Horizontal translation (gravity perpendicular to the streamwise direction): $t_f$ decreases monotonically with increasingly negative $\mathit{Ri}$, as buoyancy destabilises the lower wake, suppresses the lower spiral, and accelerates rear melting.}
    }
        \label{fig:ri-tf}
    \end{figure}

Finally, we compare the survival time of the melting sphere under different buoyancy conditions, with the initial Reynolds number fixed at $\mathit{Re}_0=200$. Figure~\ref{fig:ri-tf}(\textit{a}) shows the case of vertical translation. For moderate values $-0.2 \leq \mathit{Ri} \leq 0.3$, the complete melting time $t_f$ increases monotonically with more positive $\mathit{Ri}$, reflecting a reduced mean melting rate $\bar{v}_\mathit{\Gamma}$. This trend is consistent with the shape effect reported by \citet{yang2024shape}, who showed that cup-cap morphologies melt faster than slightly prolate ones at $\mathit{Re}_0 \sim \mathcal{O}(100)$. In our simulations, assisting buoyancy ($\mathit{Ri}>0$) delays separation and promotes a prolate morphology with lower melt rates, whereas opposing buoyancy ($\mathit{Ri}<0$) strengthens recirculation, producing a cup-cap shape with faster decay (see figure~\ref{fig:ri-slice}). 
\rea{At the extremes, deviations from this monotonic trend appear. For $\mathit{Ri} > 0.3$, suppression of the wake bubble and the progressively tighter attached flow enhance rear-side melting, thereby reducing $t_f$. For $\mathit{Ri} \leq -0.5$, strong disordered wake dynamics drive the body toward a more rounded morphology with a smaller aspect ratio (see figure~\ref{fig:ri-slice}(\textit{b})), which slows the overall melting. As $\mathit{Ri}$ decreases further below $-0.5$, the aspect ratio changes only slightly, and $t_f$ reaches a plateau.}
Figure~\ref{fig:ri-tf}(\textit{b}) shows the case of horizontal translation. Here, $t_f$ decreases monotonically with more negative $\mathit{Ri}$. The downward buoyancy plume destabilises the lower boundary layer, narrowing the lower spiral, and drives warmer liquid into the wake, thereby enhancing melting of the lower rear surface and accelerating overall decay.

\rec{
The analysis of buoyancy effects presented thus far has relied on the assumption of a linear density–temperature relation. However, as noted earlier, the ice–water system exhibits a pronounced density anomaly, for which a quadratic relation is required to capture the local maximum density at $4\,^\circ\mathrm{C}$. In appendix \ref{sec:app}, we assess how adopting this quadratic model qualitatively modifies the melting dynamics. The key observations can be summarised as follows.
}

\rec{
First, when the ambient water is at $20\,^\circ\mathrm{C}$, the cold fluid released from the melting interface must pass through $4\,^\circ\mathrm{C}$, where the density peaks within the thermal boundary layer. Nevertheless, the overall flow remains dominated by sinking cold water, consistent with the behaviour reported by \cite{weadyAnomalousConvectiveFlows2022} for warmer ambient conditions such as $T_\infty = 8^\circ\mathrm{C}$. In this regime, the density anomaly does not fundamentally alter the flow morphology; instead, it primarily shifts the separation position and modifies the size of the recirculation region. Consequently, the qualitative conclusions drawn from the linear density–temperature model remain valid for $20\,^\circ\mathrm{C}$ ambient water, as is typical for fluids without density anomalies.
}

\rec{
Second, as the ambient temperature decreases and approaches $4\,^\circ\mathrm{C}$, the anomalous rising buoyancy associated with cold fluid near the melting interface gradually overtakes the sinking component. Revisiting the upward-translation configuration in figure~\ref{fig:ri-melting-processes}(\textit{a}), this increasing dominance of anomalous upward motion leads to earlier flow separation and the formation of a larger recirculation zone. A comparable transition of flow regimes can be reproduced by varying $\mathit{Ri}$ from $0.5$ to $-0.5$, which modifies the buoyancy forcing through a similar underlying mechanism.
}

\section{Conclusion}\label{sec:conclusion}

We have performed three-dimensional sharp-interface simulations to investigate the melting of a sphere translating in a warmer liquid, systematically examining the effects of initial Reynolds number ($\mathit{Re}_0$), Stefan number ($\mathit{St}$), and Richardson number ($\mathit{Ri}$). The simulations resolve the coupled evolution of wake dynamics and interface morphology, providing a unified framework for forced and mixed-convective melting.

In the absence of buoyancy ($\mathit{Ri}\approx 0$), four distinct regimes emerge: (i) an \textit{axi-symmetric regime} ($\mathit{Re}_0<212$) with a rounded front and vertically planar rear; (ii) a \textit{steady-planar-symmetric regime} ($212<\mathit{Re}_0<273$) where the rear interface inclines; (iii) a \textit{periodic-planar-symmetric regime} ($273<\mathit{Re}_0<355$) in which vortex shedding gradually weakens as $\mathit{Re}_e$ decreases but the planar rear persists; and (iv) a \textit{chaotic regime} ($\mathit{Re}_0>355$) characterized by fluctuating stagnation points and a rounded rear. Across all regimes, the interfacial melting rate exhibits a robust tendency toward spatial homogenization over time, a feature most evident at high $\mathit{Re}_0$. Besides, classical volume-loss scaling systematically over-predicts melting in the planar-rear regimes, but incorporating an aspect-ratio-dependent correction yields a new predictive model that quantitatively captures the time evolution of the remaining volume. Hydrodynamic loads mirror this coupling: at low $\mathit{Re}_0$, drag follows rigid-sphere correlations, while at higher $\mathit{Re}_0$ it is amplified by planar rear formation; lift arises only in symmetry-broken regimes and reverses as the inclined rear plane evolves; torque develops at later stages and acts to reorient the rear surface toward vertical, in agreement with experimental observations of free spheres.

When buoyancy is introduced ($\mathit{Ri}\neq 0$), mixed convection reorganises both wake dynamics and interfacial evolution. For vertical motion, assisting buoyancy ($\mathit{Ri}>0$) stabilises the boundary layer and delays separation, while opposing buoyancy ($\mathit{Ri}<0$) enhances recirculation and accentuates planar rears. For horizontal motion, buoyancy plumes destabilise the lower boundary layer while stabilising the upper one, producing tilted rears and asymmetric wakes even below the first bifurcation threshold.

\rea{Furthermore, the present findings provide valuable physical insight into the as-yet unexplored melting dynamics of freely moving solid sphere. The path instabilities of non-melting spheres have been extensively characterized in previous studies; notably, \citet{augustePathOscillationsEnhanced2018} identified a sequence of rising trajectories, given those vertical, oblique, zigzagging, and chaotic, as the dimensionless Archimedes number $\mathit{Ar}_c = V_g D_0 / \nu$ (with $V_g$ the gravitational velocity) increases from $150$ to $700$. These canonical paths bear strong correspondence to the flow regimes reported here as $\mathit{Re}_0$ increases: the vertical trajectory parallels the axi-symmetric regime dominated by drag, the oblique path corresponds to the steady planar-symmetric regime where lift arises, the zigzagging motion reflects the periodic planar-symmetric regime characterized by vortex shedding, and the chaotic trajectory aligns with the fully three-dimensional chaotic regime. Extrapolating these path-vortex correlations, one may anticipate that freely rising melting spheres would exhibit analogous behavioural transitions. Spheres following vertical trajectories are expected to preserve this state until disappearance, maintaining a rounded front and a planar rear interface. For those undergoing steady or periodic oblique motions, the conclusions of the present torque and lift analyses suggest a tendency toward reorientation: the torque acts to align the rear interface perpendicular to the incident flow, while the decreasing effective Reynolds number progressively damps wake asymmetry and weakens lateral lift. As a result, such spheres are predicted to gradually abandon non-vertical motion and revert to a vertical ascent prior to complete melting. At higher $\mathit{Re}_0$, melting spheres initially displaying chaotic paths are unlikely to sustain persistent three-dimensional instability. Although $\mathit{Re}_0$ decreases during melting, the torque-induced reorientation provides a negative feedback that suppresses wake asymmetry, thereby attenuating chaotic oscillations. Consequently, these bodies are expected to evolve toward quasi-vertical trajectories as melting proceeds.  It should be emphasized that these predictions are extrapolated from simulations of fixed spheres. A definitive understanding of freely moving melting spheres, involving the coupled interaction between fluid flow, phase change, and rigid-body motion, requires fully coupled simulations with six degrees of freedom, which is rather a challenge that forms a natural extension of the present work.}

Taken together, these findings establish a comprehensive regime map for the melting of spheres, clarifying how interface evolution and wake dynamics co-organise under forced and mixed convection. The study advances predictive capability for both heat transfer and hydrodynamic loads, introduces a new scaling framework for volume evolution at finite Reynolds number, and reveals how buoyancy alters symmetry-breaking thresholds. Beyond their fundamental interest, these results provide a foundation for modelling particle melting in geophysical, industrial, and environmental flows.

\section{Acknowledgement}

The authors gratefully acknowledge the support of the National Key R$\&$D Program of China under grants number 2023YFA1011000, that of the NSFC under grants numbers 12222208, 12588201, 124B2046 and 12472256 and that of the ``the Fundamental Research Funds for the Central Universities'' under grants number xzy022024050.

\appendix

\section{\rec{Effects of linear and quadratic density–temperature relations on the melting dynamics}}
\label{sec:app}

\begin{figure}
    \centering
    \includegraphics[width=\textwidth]{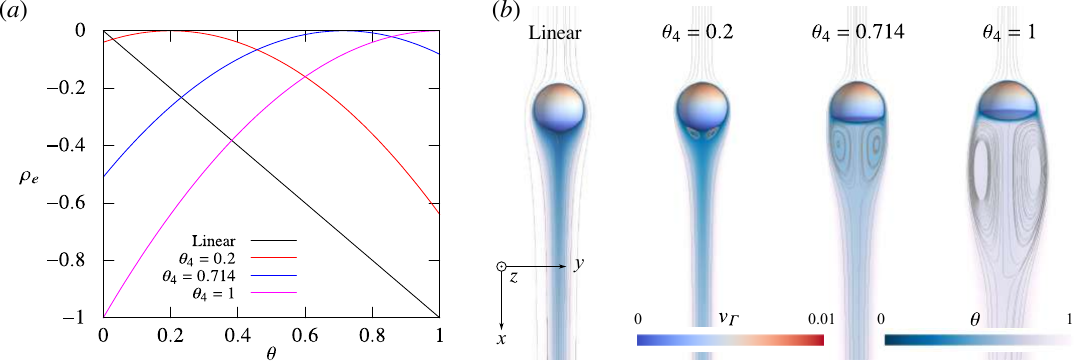}
    \caption{
\rec{(\textit{a}) Effective density $\rho_e = -(\theta - \theta_4)^2$ as a function of dimensionless temperature $\theta$ for $\theta_4 = 0.2$, $0.714$, and $1$, corresponding to $T_\infty = 20$, $5.6$, and $4\,^\circ\mathrm{C}$, respectively. The linear reference $\rho_e = -\theta$ is shown for comparison.  
(\textit{b}) Influence of $\theta_4$ on the melting dynamics of an upward-translating sphere at $\mathit{Re}_0=200$ and $\mathit{Ri}=0.5$ (see \S~\ref{sec:gravity-perpendicular}). As $\theta_4$ decreases, rising cold water intensifies buoyant circulation, advancing separation and enlarging the recirculation zone. The linear case is included as a reference.}
}
    \label{fig:a2}
\end{figure}

\rec{The main body of this study (\S~\ref{sec:buoyancy}) assumes a linear dependence of liquid density on temperature. However, for water and similar fluids exhibiting a density anomaly, a quadratic relation provides a more realistic description, particularly for the ice–water system \citep{weadyAnomalousConvectiveFlows2022,yang2024circular,wang2021growth}. This appendix examines how such a quadratic model modifies the melting dynamics. Under this assumption, the density–temperature relation is expressed as}
\begin{equation}\label{eq:a1}
    \rec{\rho(T) = \rho_4 \left[ 1 - \beta_{\mathit{ano}}(T - 4\,^\circ \mathrm{C})^2 \right]},
\end{equation}
\rec{where $\rho_4$ denotes the density at $4\,^\circ \mathrm{C}$ and $\beta_{\mathit{ano}}$ is the quadratic anomaly coefficient. Within the Boussinesq approximation, the buoyancy term in the non-dimensional momentum equation (\ref{eq:ns1}) becomes
$-\mathit{Ri}_{\mathit{ano}}(\theta - \theta_4)^2 \bm{e}_i$,
in contrast to the linear form $-\mathit{Ri}\,\theta\,\bm{e}_i$. The corresponding anomalous Richardson number is
$\mathit{Ri}_{\mathit{ano}} = g\beta_{\mathit{ano}}(T_\infty - T_0)^2 D_0/U_\infty^2$,
and $\theta_4 = (4\,^\circ\mathrm{C} - T_0)/(T_\infty - T_0)$ denotes the dimensionless temperature associated with $4\,^\circ\mathrm{C}$. Figure~\ref{fig:a2}(\textit{a}) shows representative profiles of the effective density $\rho_e = -(\theta - \theta_4)^2$ for several $\theta_4$, which serve as a useful reference for interpreting the following results.}

\rec{Through combined experiments and simulations of natural-convection-driven melting, \citet{weadyAnomalousConvectiveFlows2022} identified three characteristic regimes depending on the ambient temperature $T_\infty$. For $T_\infty = 8\,^\circ\mathrm{C}$, the flow is dominated by sinking cold water; for $T_\infty = 4\,^\circ\mathrm{C}$, an upward motion emerges due to the density anomaly; and at intermediate conditions ($T_\infty \approx 5.6\,^\circ\mathrm{C}$), a shear-driven circulation develops—rising colder water ($T<4\,^\circ\mathrm{C}$) near the interface and sinking slightly warmer water ($4<T<5.6\,^\circ\mathrm{C}$) farther away. Motivated by these observations, we explore the effects of the quadratic relation in two stages. First, we consider a warm ambient flow at $T_\infty = 20\,^\circ\mathrm{C}$, corresponding to the baseline condition of the present study. The dimensionless equivalent of $T=4\,^\circ\mathrm{C}$ is $\theta_4 = 0.2$. We compare the melting behaviour predicted by the linear and quadratic relations under identical parameters. Second, we vary the ambient temperature by selecting $T_\infty = 4$, $5.6$, and $20\,^\circ\mathrm{C}$ (i.e. $\theta_4 = 1$, $0.714$, and $0.2$, respectively) to explore how the density anomaly modulates buoyancy effects. In both stages, we fix $|\mathit{Ri}_{\mathit{ano}}| = |\mathit{Ri}| = 0.5$ and $\mathit{Re}_0 = 200$ to isolate the influence of the density–temperature model.}

\begin{figure}
    \centering
    \includegraphics[width=\textwidth]{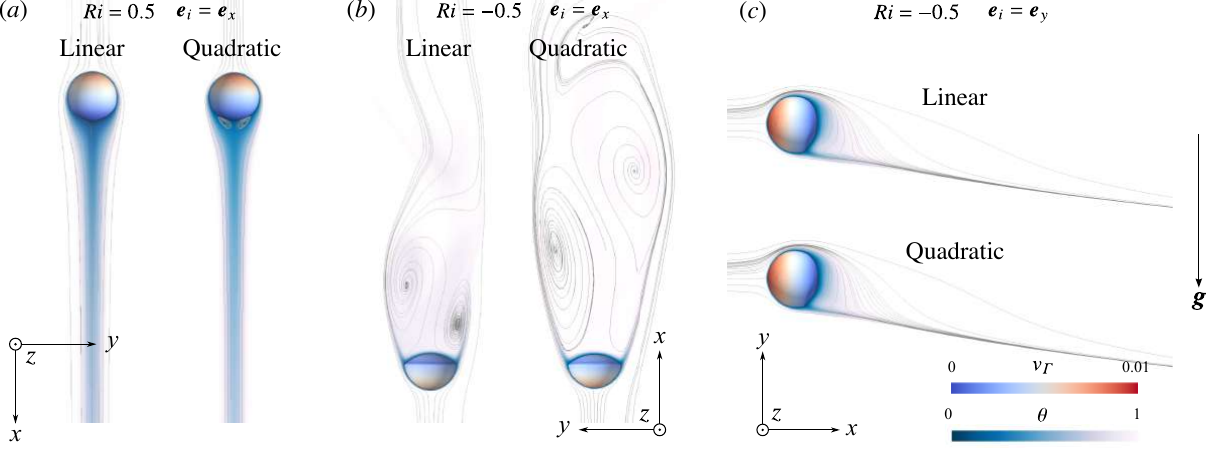}
    \caption{
    \rec{Comparison between the linear density–temperature relation and the quadratic anomaly model (\ref{eq:a1}) for $\theta_4 = 0.2$ ($T_\infty = 20\,^\circ\mathrm{C}$), with $|\mathit{Ri}_{\mathit{ano}}| = |\mathit{Ri}| = 0.5$ and $\mathit{Re}_0 = 200$. Panels (\textit{a--c}) correspond to the configurations discussed in \S~\ref{sec:buoyancy}: (\textit{a}) upward translation, (\textit{b}) downward translation, and (\textit{c}) horizontal translation. Snapshots are taken at $t=50$ for panels (\textit{a,b}) and at $t=40$ for panel (\textit{c}). The anomaly enhances buoyant motion in vertical configurations but has limited influence in horizontal translation.}
}
    \label{fig:a1}
\end{figure}

\rec{Figure~\ref{fig:a1} compares the melting behaviour obtained using the linear and quadratic relations for the buoyancy configurations discussed in \S~\ref{sec:buoyancy}. When the anomaly is included ($\theta_4=0.2$, $T_\infty=20\,^\circ\mathrm{C}$), several clear differences arise. In upward translation (panel (\textit{a})), the anomaly intensifies the upward buoyant motion of the freshly melted cold water, generating a small rear recirculation bubble, whereas the flow remains attached under the linear assumption. For downward translation (panel (\textit{b})), separation occurs earlier and the wake enlarges when the anomaly is included. This behaviour is consistent with the density profiles in figure~\ref{fig:a2}(\textit{a}), where $\theta_4=0.2$ yields a steeper $-\mathrm{d}\rho_e/\mathrm{d}\theta$ near $\theta\to1$, implying a stronger downward buoyant acceleration within the thermal boundary layer. For horizontal translation (panel (\textit{c})), the anomaly has a comparatively minor effect. In this configuration, the downward motion of the melt dominates the wake dynamics, and separation occurs on the lower side; the flow orientation is thus governed primarily by the sign and magnitude of $\mathit{Ri}$ (or $\mathit{Ri}_{\mathit{ano}}$) rather than by the specific density–temperature law.}

\rec{We next examine the influence of varying $\theta_4 = 0.2$, $0.714$, and $1$, corresponding to decreasing $T_\infty$ toward $4\,^\circ\mathrm{C}$. As shown in figure~\ref{fig:a2}(\textit{a}), decreasing $\theta_4$ produces a transition from sinking-dominated buoyancy, through a mixed regime, to rising-dominated buoyancy driven by the anomalous density inversion near $4\,^\circ\mathrm{C}$. The corresponding melting dynamics for the upward-translation configuration (\S~\ref{sec:gravity-perpendicular}) are shown in figure~\ref{fig:a2}(\textit{b}). As the effect of rising cold water intensifies, flow separation advances upstream and the rear recirculation region expands. The resulting sequence of wake transitions mirrors those obtained by varying $\mathit{Ri}$ from $0.5$ to $-0.5$ under the linear law (figures~\ref{fig:ri-slice} and \ref{fig:ri-melting-processes}), indicating that both the linear and anomalous buoyancy mechanisms act through comparable physical processes.}

\bibliographystyle{jfm}
\bibliography{jfm}

\end{document}